\documentclass[useAMS,usenatbib]{mn2e}
\usepackage{amsmath}
\usepackage{amssymb}
\usepackage{threeparttable}
\usepackage{subfigure}
\usepackage{cases}
\usepackage{color}
\usepackage{graphicx}
\usepackage{hyperref}
\usepackage[figure]{hypcap}
\usepackage{tikz}

\hypersetup{colorlinks, citecolor=blue, pdfduplex=DuplexFlipLongEdge}
\hypersetup{pdftitle=HERO paper, pdfauthor=Yucong Zhu, pdfsubject=Astrophysics}

\title[HEROIC: 3D GR Radiative Code Transfer with Comptonization]
  {HEROIC: 3D General Relativistic Radiative Postprocessor with Comptonization for Black Hole Accretion Discs}
\author[R. Narayan, Y. Zhu, D. Psaltis, \& A. S\c{a}dowski]
  {Ramesh Narayan$^1$\thanks{E-mail: \hbox{rnarayan@cfa.harvard.edu~(RN);} \hbox{dpsaltis@email.arizona.edu~(DP);} \hbox{asadowsk@mit.edu~(AS)}}
 Yucong Zhu$^1$,
 Dimitrios Psaltis$^2$\footnotemark[1],
 Aleksander S\c{a}dowski$^3$\footnotemark[1]
\\
  $^1$Harvard-Smithsonian Center for Astrophysics, 60 Garden Street, Cambridge, MA 02138, USA\\
  $^2$University of Arizona, 933 N. Cherry Ave, Tucson, AZ 85721, USA\\
  $^3$MIT Kavli Instiatute for Astrophysics and Space Research, 77 Massachusetts Ave, Cambridge, MA 02139, USA}

\begin{document}

\maketitle

%====================================================================================================

\begin{abstract}

We describe HEROIC, an upgraded version of the relativistic radiative
post-processor code HERO described in a previous paper, but which now
Includes Comptonization.  HEROIC models Comptonization via the
Kompaneets equation, using a quadratic approximation for the source
function in the short characteristics radiation solver. It employs a
simple form of accelerated lambda iteration to handle regions of high
scattering opacity. In addition to solving for the radiation field,
HEROIC also solves for the gas temperature by applying the condition
of radiative equilibrium.  We present benchmarks and tests of the
Comptonization module in HEROIC with simple 1D and 3D scattering
problems. We also test the ability of the code to handle various
relativistic effects using model atmospheres and accretion flows in a
black hole space-time.  We present two applications of HEROIC to
general relativistic MHD simulations of accretion discs. One
application is to a thin accretion disc around a black hole. We find
that the gas below the photosphere in the multi-dimensional HEROIC
solution is nearly isothermal, quite different from previous solutions
based on 1D plane parallel atmospheres. The second application is to a
geometrically thick radiation-dominated accretion disc accreting at 11
times the Eddington rate. The multi-dimensional HEROIC solution shows
that, for observers who are on axis and look down the polar funnel,
the isotropic equivalent luminosity could be more than ten times the
Eddington limit, even though the spectrum might still look thermal and
show no signs of relativistic beaming.

\end{abstract}

%====================================================================================================
%====================================================================================================

\section{Introduction}\label{sec:intro}

Comptonization plays a crucial role in determining the high-energy
emission properties of a variety of astrophysical objects, e.g., X-ray
emission from X-ray binaries \citep{white88,ponman90} and from active
and quiescent galactic nuclei \citep[see, e.g., ][and references
  therein]{haardt91,yuan_narayan14}, nonthermal spectral properties of
galactic microquasars \citep{mcclintockremillard06,done07}, cooling
rates and decay timescales of X-ray bursters \citep{joss77},
scattering by hot plasma clouds in intra-cluster media
\citep{prokhorov10}.  The ubiquity of hot ionized gas in astrophysical
settings motivates the need for accurate treatment and modelling of
the Compton scattering process.

In the case of black hole accretion discs, cold seed photons emitted
from a thermal disc are upscattered by hot coronal electrons
\citep{sunyaevtitarchuk80}, producing a whole host of spectral shapes
in X-rays such as power-laws and Compton humps.  Many systems,
especially those in the low-hard spectral state, are observed with a
dominant Compton component \citep{grove88,gierlinski97,
  mcclintockremillard06}, where interpretation of the data requires
accurate modeling of the power-law tail.  This is especially true in
the case of reflection line modeling for black hole systems
\citep{tanaka95,reynolds14}, where slight errors in the continuum can
lead to systematic biases in the derived black hole parameters
\citep{haardt93}.
%\citep{weaver98,miller06,kolehmainen10,kolehmainen14}.

Accurate analytic models of the Comptonized spectrum have been worked
out during the last several decades for various regimes such as for
optically thick, homogeneous 1D and 3D media
\citep{sunyaevtitarchuk80}, at the limit of low electron energies
\citep{haardt93}, at the relativistic limit for optically thin
media~\citep{coppi91}, and for bulk Comptonization~\citep[see][and
  references therein]{turolla02}.  However, the complex nature of
time-dependent simulations of accretion flows around black holes
precludes the use of analytic models, motivating the development of
numerical Comptonization schemes.

Monte-Carlo based methods have been a popular approach to the problem
\citep{pozdnyakov83,gorecki84,stern95,dolence09,kawashima12,schnittman13,moscibrodzka14}
owing to the ease with which the technique can handle relativistic
geometries and the complex angle and frequency dependent scattering
process.  The biggest drawback of the Monte-Carlo approach is that
photon statistics limit the accuracy of the computations.  This
problem is worst in the limit of photon energies much higher than the
injection energy, where there is a dearth of photons and hence poor
photon statistics \footnote{The use of an energy-weighted scattering
  kernel is one work-around for the photon-starvation problem.}.
Monte Carlo methods also suffer in the limit of high optical depths --
here, the full photon diffusion process is computationally very
taxing, which in practice restricts MC-based codes to problems with
only moderate optical depths ($\tau \lesssim 10$).

Another approach is to discretize the problem and numerically solve
the radiative transfer problem for some preset fixed geometry (e.g.,
\textsc{compPS}: \citealt{poutanen96} for moderate optical depths;
\textsc{compTT}: \citealt{titarchuk95} for accretion discs;
\textsc{TLUSTY}: for 1D atmospheres \citealt{hubeny01};
\citealt{zane96} and \citealt{psaltis01} for bulk Comptonization).
The main advantage of this approach is that it can easily handle
optically thick problems via a Kompaneets operator approach, which
uses a diffusion approximation to handle the nonrelativistic
Comptonization problem.  This approach is particularly amenable to the
short characteristics fixed-grid framework of HERO (Hybrid Evaluator
for Radiative Objects), a 3D GR radiation postprocessor code which we
described in a previous paper \citep[][hereafter Paper 1]{zhu15}. Here
we implement Comptonization in HERO and update the name of the code to
HEROIC (HERO Including Comptonization).  The primary advance in our
work is that we introduce a self-consistent relativistic radiation
module for the 3D Comptonization problem of hot accretion flows around
black holes.

Regardless of the approach taken for solving the Compton problem, a
final raytracing calculation is needed to connect the result to the
actual spectral observations of astrophysical systems.  Typically,
geodesic paths are traced backwards from a distant observation plane
until they hit the accretion flow (see, e.g., \citealt{rauch94,
  broderick03, dovciak04, dexter09, kulkarni11, psaltis12, zhu12}).
This yields a transfer function that allows one to map the local
Comptonized disc emission to the spectrum as measured by the distant
observer.  In cases of high scattering optical depths, it is crucial
for raytracing methods to resolve the complete nonlocal structure of
the scattered radiation field \citep{schnittman13}.  For Monte
Carlo-based methods, this translates to a more computationally
expensive ``emitter-to-observer'' paradigm since this is how the
photon diffusion process works in nature (see \citealt{laor90,
  kojima91, dolence09, schnittman13} for a few recent codes that
follow this philosophy).  Grid based methods instead require a fully
3D treatment of the radiative problem accounting for all the nonlocal
scattering terms in the emissivity profile.  In the case of HEROIC,
this is achieved by solving for the complete 3D scattered radiation
field everywhere around and inside the disc before the raytracing
process is initiated.
  
The organization of this paper is as follows. In \S\ref{sec:methods}
we describe the methodology used by HEROIC, focusing in particular on
Comptonization; specifically, we explain how the ray evolution
equation works in the presence of Compton scattering and how we solve
it using a Kompaneets-based approach. We also describe how we solve
self-consistently for the temperature of the radiating medium.  This
is followed in \S\ref{sec:NumericalTests} with a series of 1D and 3D
benchmark tests to verify the correct operation of the code.  Then, in
\S\ref{sec:disc} we present two applications of HEROIC to data
obtained with general relativistic magnetohydrodynamic (GRMHD) disc
simulations.  Finally, in \S\ref{sec:discussion} we conclude with a
discussion.

\section{Methods}\label{sec:methods}

HEROIC solves the radiative transfer equation iteratively to obtain a
steady state solution for the radiation intensity $I_\nu$ as a
function of position ${\bf r}$, frequency $\nu$ and ray direction
${\bf n}$.  If the problem requires it, HEROIC also applies the
condition of radiative equilibrium to solve for the temperature of the
gas in each grid cell.  Other fluid quantities, specifically, the
density $\rho$ and four-velocity $u^\mu$, are kept fixed. Thus, HEROIC
solves the radiative transfer problem, but does not deal with the
dynamics of the fluid. The latter should be specified as part of the
initial setup, and would usually be obtained from a general
relativistic radiation hydrodynamics or MHD simulation (e.g.,
\citealt{sadowski14,mckinney14,fragile14,sadowski15b,takahashi_ohsuga15}).
HEROIC assumes that the system is time-steady. Therefore, it is
best-suited for objects in steady state. If the code is applied to
time-varying systems, then effectively one makes the ``fast-light''
approximation, i.e., one neglects time-travel delays between different
regions of the source.

In a typical black hole accretion disc application, one assumes
axisymmetry and solves the problem on a two-dimensional spatial grid in
Boyer-Lindquist (polar) coordinates $r$-$\theta$ in the Kerr
space-time of the black hole. At each grid point, the radiation field
is decomposed over a uniform grid of angles, typically $N_A=80$ angles,
covering the full $4\pi$ steradians, and over a grid of frequencies,
typically 10 frequencies per decade distributed uniformly in
$\log\nu$. The intensities $I_\nu$, the mean intensity,
\begin{equation}
J_\nu = \frac{1}{4\pi} \int I_\nu({\bf n}) \,d\Omega, \label{eq:jnu}
\end{equation}
the opacity, the temperature, etc., are all described in the local
comoving frame of the fluid.  However, ray geodesics are best computed
in the fixed spatial grid of the ``lab'' (Boyer-Lindquist) frame.
Since we are using a form of the radiative transfer equation that is
relativistically invariant, transforming quantities from one frame to
the other is straightforward.

\subsection{Basic Equations}\label{sec:equation}

%\DP{I'm a little concerned with the fact that, even though the
%  transfer equation in HEROIC is solved in the proper spacetime with
%  the appropriate coordinates, the following two equations seem to
%  indicate that they are done in flat spacetime. For example, ds, in
%  equation (3) is a proper distance, which involves all the metric
%  elements, depending on the orientation of the ray. Shouldn't we put
%  all that in, especially to quantify how we deal with large cells
%  along which the metric elements are not constant? This will also
%  allow us to put in the refshift factors, etc, so that when we talk
%  about tests later on, it will become obvious which terms contribute
%  and why.} 

%\RN{Dimitrios, this is a good point but I am not sure how to explain
%  it all without becoming overly technical. Also, I am getting tired
%  of this paper and want to get rid of it! There is whole subsection
%  in the HERO paper in which Yucong discusses the relativistic version
%  of the transfer equation. I have included a footnote here.}

HEROIC solves the radiative transfer problem iteratively using the
method of characteristics, as described in more detail in Paper 1.

The radiative transfer equation states that the intensity $I_\nu$ of a
ray evolves along the ray trajectory according to
\begin{equation}
\frac{dI_\nu}{d\tau_\nu}=-I_\nu(\tau_\nu) + S_\nu(\tau_\nu),\label{eq:radTrans}
\end{equation}
where $\tau_\nu$ is the optical depth at frequency $\nu$ measured
along the ray and is given by
\begin{equation}
d\tau_\nu = (\kappa_\nu + \sigma_\nu) ds,
\end{equation}
where $\kappa_\nu$ and $\sigma_\nu$ are the absorption and scattering
coefficients and $s$ is distance along the ray.\footnote{Although, for
  simplicity, we have written the transfer equation here in
  non-relativistic notation, everything is done in a relativistically
  covariant form within HERO and HEROIC (see Paper 1 for more
  details).}  The quantity $S_\nu$ is the source function, which
governs the rate at which energy is introduced into the beam,
accounting for both intrinsic thermal emission and scattering.  The
formal solution of the radiative transfer equation (\ref{eq:radTrans})
for the intensity $I_\nu(\tau_{\nu,2})$ at some location labeled by
optical depth $\tau_{\nu,2}$ can be written in terms of the intensity
$I_\nu(\tau_{\nu,1})$ at another location $\tau_{\nu,1} <
\tau_{\nu,2}$ (i.e., located at some earlier point along the ray
trajectory) and the source function between the two locations as
follows:
\begin{equation}\label{eq:formal}
I_\nu(\tau_{\nu,2}) = I_\nu(\tau_{\nu,1}) e^{\tau_{\nu,1} - \tau_{\nu,2}}
+ \int_{\tau_{\nu,1}}^{\tau_{\nu,2}} S_\nu(\tau_{\nu}) e^{\tau_{\nu} - \tau_{\nu,2}} d\tau_\nu.
\end{equation}
Paper 1 describes at some length how the code HERO uses the above
formal solution to iteratively solve for the radiation field over the
entire grid.

The inclusion of Comptonization in HEROIC introduces two changes
relative to the discussion given in Paper 1.  First, the source
function now takes the form
\begin{equation}
S_\nu = \epsilon_\nu B_\nu + (1-\epsilon_\nu) J_{\nu,\rm Compt},
\end{equation}
where $\epsilon_\nu$ is the ratio of absorption to total opacity,
\begin{equation}
\epsilon_\nu = \frac{\kappa_\nu}{\kappa_\nu + \sigma_\nu},
\end{equation}
$B_\nu(\tau_\nu)$ is the Planck function corresponding to the local
temperature at location $\tau_\nu$, and $J_{\nu,\rm Compt}$, which in
Paper 1 was simply equal to the mean intensity $J_\nu$, now depends on
the details of Compton-scattering. Specifically, if radiation with a
local mean intensity distribution $J_\nu$ scatters once off the hot
electrons in the Comptonizing medium, then $J_{\nu,\rm Compt}$ is the
resulting intensity distribution.

For notational convenience later on, we rewrite the source function in
terms of the uncomptonized $J_\nu$ by introducing an ``amplification
factor'' $a_\nu$,
\begin{equation}\label{eq:snucompt}
S_\nu = \epsilon_\nu B_\nu + (1 + a_\nu) J_\nu,
\end{equation}
where we have absorbed all the complexities of Comptonization into
$a_\nu$. The latter is defined by the relation
\begin{equation}\label{eq:jnucompt}
J_{\nu,\rm Compt} \equiv A_\nu J_\nu \equiv \frac{1+a_\nu}{1-\epsilon_\nu} J_\nu,
\end{equation}
where $A_\nu$ describes the boost factor in the radiation field at
frequency $\nu$ due to Compton scattering.  Note that Compton
scattering mixes radiation at different frequencies, so the quantities
$A_\nu$ and $a_\nu$ are functions not just of the post-scattering
frequency $\nu$ but also of the pre-scattering frequency. Equations
(\ref{eq:snucompt}) and (\ref{eq:jnucompt}) are thus valid only if
$a_\nu$ and $A_\nu$ are defined for a specific pre-scattering
intensity distribution $J_\nu$. This is not a limitation for our
purposes since these quantities are constantly recomputed based on the
current solution as the iterations in HEROIC proceed.

The second difference due to Comptonization is that, when evaluating
the integral in equation (\ref{eq:formal}), we find it necessary to
expand $S_\nu(\tau_\nu)$ versus $\tau_\nu$ up to at least the
quadratic term:
\begin{equation}\label{eq:snuseries}
S_\nu(\tau_\nu) = S_\nu(\tau_{\nu,2}) + S'_\nu (\tau_\nu -
\tau_{\nu,2}) + \frac{1}{2} S''_\nu (\tau_\nu - \tau_{\nu,2})^2 +
\cdots.
\end{equation} 
Quadratic order is helpful for any problem that has a source of
heating or cooling which results in a transfer of energy from gas to
radiation or vice versa. But it is particularly important in the case
of Comptonization under very optically thick conditions, as we have
found during the tests discussed in \S\S\ref{sec:planecompt},
\ref{sec:sphericalcompt}. In the work described in this paper, we have
truncated the series at the quadratic term (Paper 1 stopped at the
linear term). The evaluation of the coefficients $S'_\nu$ and
$S''_\nu$ in equation (\ref{eq:snuseries}) is discussed in
\S\ref{sec:quadratic}.

Notice that the radiative transfer problem involves an intimate
coupling between the intensities and the source function.  Ray
intensities are computed from the spatially varying source function
via equation (\ref{eq:formal}).  However, the source function itself
depends on the radiation field through $J_{\nu,{\rm Compt}}$, which
depends on $J_\nu$ and $a_\nu(J_\nu)$, and is ultimately determined by
the local intensities.  Solving in parallel for the temperature only
adds to the complexity. We use the lambda iteration technique with
acceleration (\S\ref{sec:ALI}) to solve the radiative transfer part of
the problem, and have developed other techniques to solve for the
temperatures (\S\ref{sec:temperature_solve}).

\subsection{Compton Boost Factor}\label{sec:KompRay}

The Compton boost factor $A_\nu$ is locally defined and computed in
each spatial cell. It describes the effect of Compton scattering by
electrons with the temperature $T$ of this cell\footnote{We do not
  distinguish between the electron temperature $T_e$ and the gas
  temperature $T_{\rm gas}$. We refer to both as $T$.} on the mean
radiation intensity $J_\nu$ in the cell. Both $T$ and $J_\nu$ change
from one iteration to the next, so $A_\nu$ is computed afresh in each
iteration.

HEROIC computes $A_\nu$ by solving for the evolution of the photon
number density,
\begin{equation}\label{eq:nnu}
n_\nu = \left(\frac{c^2}{2h}\right) \frac{J_\nu}{\nu^3} ,
\end{equation}
as a result of scattering. The evolution is computed via the
Kompaneets equation,
\begin{equation}
\frac{\partial n}{\partial t_{\text{scatt}}} = f(\theta_e)
\frac{1}{x^2}\frac{\partial}{\partial x}\left(x^4\left[\frac{\partial
    n}{\partial x} + n(n+1)\right]\right), \label{eq:Kompaneets}
\end{equation}
where $t_{\text{scatt}}$ measures the characteristic time of the
system in units of the number of scattering events, $\theta_e$ is the
dimensionless electron temperature,
\begin{equation}\label{eq:thetae}
\theta_e = \frac{kT}{m_ec^2},
\end{equation}
$x$ is the dimensionless frequency,
\begin{equation}\label{eq:x}
x=h\nu/kT,
\end{equation}
and the function $f(\theta_e) = \theta_e$ in the limit of
non-relativistic temperatures, as originally considered by Kompaneets,
but can be approximated at relativistic temperatures by
\begin{equation}\label{eq:relcompt}
f(\theta_e) \approx  \frac{(1+3.683\theta_e+4\theta_e^2)}{(1+\theta_e)}\,
\theta_e,
\end{equation}
as discussed in the Appendix.  

We are interested in the radiation field $J_{\nu,\rm Compt}$ after a
single scattering event.  This corresponds to starting with an initial
photon number density $n_{\nu,\rm initial}$ given by equation
(\ref{eq:nnu}) for the current mean intensity $J_\nu$, solving
Equation (\ref{eq:Kompaneets}) for the photon distribution $n_{\nu,\rm
  final}$ after a time $\delta t_{\text{scatt}} = 1$, and computing
from this the Compton boost factor,
\begin{equation}
A_{\nu} = \frac{J_{\nu,\rm final}}{J_\nu} = \frac{n_{\nu,\rm
    final}}{n_{\nu,\rm initial}}.
\end{equation}

Due to the stiff nature of equation (\ref{eq:Kompaneets}), we solve
the partial differential equation using the approach described in
\citet{changcooper70}.  The system is discretized along a logarithmic
frequency grid and the photon fluxes in neighboring frequency bins
are chosen such that the expected quasiequilibrium state for the given
gas temperature and total photon number remains stationary.  This
guarantees convergence towards the expected thermal photon
distribution, and guards against instabilities that can arise from the
stiffness of the equation.  Additionally, since the Kompaneets
equation is a diffusion equation, the coupling is only between
neighboring frequency bins, which results in a simple tridiagonal
system that is easy to invert using standard methods.  The boundary
conditions are zero photon flux at the lower and upper frequency
boundaries to ensure conservation of photon number:
\begin{equation}
\frac{\partial n}{\partial x} + n(n+1) = 0, \qquad x = x_{\rm
  min},~x_{\rm max}.
\end{equation}
The \citet{changcooper70} approach has been successfully tested and
applied in other codes \citep{pomraning73,madej89,hubeny01},
with some implementations being more sophisticated than ours due to
differences in the choice of interpolation scheme.

One detail is worth mentioning. When the scattering optical depth
$\Delta\tau_{\rm scatt}$ across a spatial cell is large, the mean
number of scatterings experienced by a photon as it moves across the
cell, $n_{\rm scatt} \sim (\Delta\tau_{\rm scatt})^2$, can be much
larger than unity. We have then found that it is better to evolve the
Kompaneets equation over a time $t_{\rm scatt} = n_{\rm scatt}$ rather
than $t_{\rm scatt} = 1$, and to correspondingly estimate $A_\nu$ from
the output of the Kompaneets equation by
\begin{equation}
A_\nu = 1 + \frac{n_{\nu,\rm final}(x) - n_{\nu,\rm
    initial}(x)}{n_{\rm scatt} \, n_{\nu,\rm initial}(x)}.
\end{equation}
The precise choice of $n_{\rm scatt}$ is not critical though we find
that it is better to scale it with $(\Delta\tau_{\rm scatt})^2$ rather
than $\Delta\tau_{\rm scatt}$. In HEROIC we use
\begin{equation}
n_{\rm scatt} = 1 + \Delta\tau_{\rm scatt} + (\Delta\tau_{\rm scatt})^2.
\end{equation}
The optical depth $\Delta\tau_{\rm scatt}$ is estimated by considering
a typical trajectory through the cell. Again the precise choice is not
important. One other detail: when the Compton $y$-parameter across the
cell, which is roughly equal to $n_{\rm scatt}\theta_e$, becomes large,
we limit $n_{\rm scatt}$ such that $y\sim$\,few.

Since the Kompaneets equation is based on a Fokker-Planck approach, it
is valid only when the change in the frequency of a photon in a single
scattering is small. This condition breaks down at large temperatures.
With $f(\theta_e)=\theta_e$, the solution deviates already at $kT_e
\approx 20$\,keV, but with the generalized $f(\theta_e)$ given in
equation~(\ref{eq:relcompt}) and discussed in the Appendix, one might
be able to go up to 100\,keV.  In fact, the equation will continue to
give smooth well-behaved solutions, and the solution will still have
the right qualitative behavior, at even higher temperatures, only
accuracy will be lost.

\subsection{Quadratic Variation of the Source Function}\label{sec:quadratic}

\begin{figure}
\begin{center}
\vspace{-1.5cm}  
\includegraphics[width=0.5\textwidth]{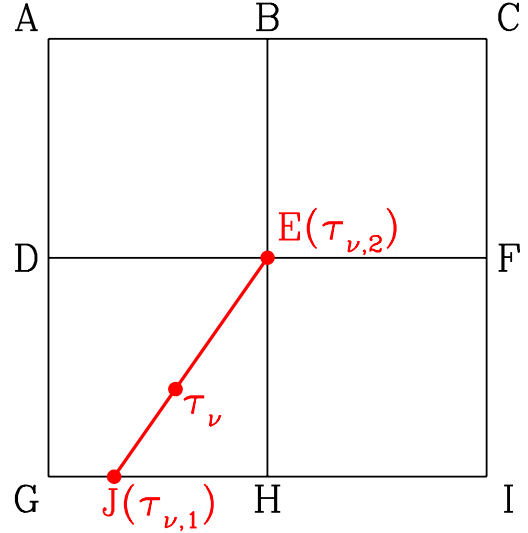}
\vspace{-2.5cm}  
\caption{Schematic plot showing the ray geometry corresponding to the
  short characteristics method used in HEROIC. The point E is the
  reference cell, and its neighboring cells are A, B, C, D, F, G, H, I
  (in 2D geometry).  Given a particular direction ${\bf n}$ along
  which one wishes to calculate the radiation intensity $I_\nu$, one
  computes a null geodesic in the backward direction (the thick red
  line) until the geodesic intersects one of the neighboring cell
  boundaries, indicated by the point J. The radiative transfer
  equation (\ref{eq:formal}) is integrated from the point J at optical
  depth $\tau_{\nu,1}$ to the point E at $\tau_{\nu,2}$. This is
  repeated for all ray directions ${\bf n}$ at E and for all
  frequencies $\nu$, thereby building up an estimate of the radiation
  field at E.}
\label{fig:short_char}
\end{center}  
\end{figure}  

In the presence of optically thick Comptonization, the radiative
transfer solution behaves correctly only if we keep terms up to the
quadratic order in the source function (\ref{eq:snuseries}).  In the
context of the short characteristics method, the point $\tau_{\nu,2}$
in equation (\ref{eq:formal}) corresponds to the particular cell where
one is interested in computing the intensity of a ray, indicated by
the point E in Figure~\ref{fig:short_char}, and $\tau_{\nu,1}$
corresponds to the point J on a neighboring cell boundary.  The source
function at E is known, and that at J is estimated by interpolating
between the known source functions at G and H. The series expansion
(\ref{eq:snuseries}) is then used to represent the source function at
intermediate points such as that labeled $\tau_\nu$ in the figure.

Given the source function at two boundary points ($\tau_{\nu,2}$,
$\tau_{\nu,1}$), one can immediately obtain a linear approximation for
$S_\nu(\tau_\nu)$ in between, and this was the approach taken in Paper
1. Obtaining a quadratic approximation is less straightforward,
especially if we do not wish to involve additional neighboring cells
in the estimate.  We use the radiative transfer equation itself to
estimate $S''_\nu$. Differentiating equation (\ref{eq:snucompt}) twice
with respect to $\tau_\nu$, we have
\begin{equation}\label{eq:d2sdtau2}
\frac{d^2 S_\nu}{d\tau_\nu^2} \approx \frac{d^2 (\epsilon_\nu
  B_\nu)}{d\tau_\nu^2} + (1+a_\nu) \frac{d^2 J_\nu}{d\tau_\nu^2},
\end{equation}
where we have ignored the variation of $a_\nu$ along the ray. To
estimate the second derivative of $J_\nu$ we proceed as follows.

Following standard definitions in radiative transer theory
\citep{mihalas78}, the zeroth angular moment of the radiation field is
$J_\nu$ as defined in eq.~(\ref{eq:jnu}), and the first and second
angular moments are
\begin{eqnarray}
{\bf H}_\nu &=& \frac{1}{4\pi} \int I_\nu({\bf n})\, {\bf n}\, d\Omega, \label{eq:hnu} \\
{\bf K}_\nu &=& \frac{1}{4\pi} \int I_\nu({\bf n})\, {\bf n}{\bf n} \,d\Omega, \label{eq:knu}
\end{eqnarray}
where ${\bf n}$ is a unit vector along the direction of a ray, and
${\bf H_\nu}$ and ${\bf K_\nu}$ are related to the flux vector and the
pressure tensor, respectively. By taking moments of the radiative
transfer equation (\ref{eq:radTrans}), one obtains
\begin{eqnarray}
{\bf \nabla_\tau \cdot  H_\nu} &=& -J_\nu + S_\nu, \label{eq:fluxRT}\\
{\bf \nabla_\tau \cdot K_\nu} &=& -H_\nu, \label{eq:pressureRT}
\end{eqnarray}
which gives the following second-order partial differential equation
for ${\bf K_\nu}$,
\begin{equation}\label{eq:nablaK}
{\bf \nabla_\tau^2 K_\nu} = J_\nu - S_\nu.
\end{equation}
Here, the subscript $\tau$ in ${\bf \nabla_\tau}$ is to indicate that
we are considering spatial gradients in terms of the optical depth
$\tau_\nu$. In regions of large optical depth, the pressure tensor is
expected to be diagonal and isotropic, so we write (this corresponds
to the Eddington closure relation):
\begin{equation}\label{eq:eddington}
{\bf K_\nu} \approx {\rm diag}(J_\nu/3).
\end{equation}
Equation (\ref{eq:nablaK}) then gives
\begin{equation}
\nabla_\tau^2 J_\nu = 3 (J_\nu - S_\nu) = -3 (\epsilon_\nu B_\nu +
a_\nu J_\nu).
\end{equation}
As our final approximation, we assume that the second derivative of
$J_\nu$ with respect to $\tau_\nu$ is isotropic, which then gives the
relation we seek,
\begin{equation}
\frac{d^2 J_\nu}{d\tau_\nu^2} \approx -\epsilon_\nu B_\nu - a_\nu
J_\nu.
\end{equation}
Substituting this in equation (\ref{eq:d2sdtau2}) and neglecting the
second derivative of $\epsilon_\nu B_\nu$, we finally obtain
\begin{equation}\label{eq:s2nu}
S''_\nu \equiv \frac{d^2 S_\nu}{d\tau_\nu^2} \approx - \epsilon_\nu
B_\nu - a_\nu S_\nu.
\end{equation}
This expression, which involves no derivatives and can be evaluated
locally in each cell, is used in HEROIC for estimating $S''_\nu$.
Once we have $S_\nu(\tau_{\nu,2})$, $S_\nu(\tau_{\nu,1})$ and
$S''_\nu(\tau_{\nu,2})$, we can immediately estimate
$S'_\nu(\tau_{\nu,2})$ by making use of equation (\ref{eq:snuseries}).

A number of approximations were made en route to deriving equation
(\ref{eq:s2nu}), but they are all harmless.  The inclusion of the
quadratic term $S''_\nu$ is important only in optically thick, highly
Comptonized regions, and in these regions we believe the terms we have
retained are the important ones. In principle, we could avoid some of
the approximations, e.g., we could retain the term $d^2(\epsilon_\nu
B_\nu) / d\tau_\nu^2$ in equation (\ref{eq:d2sdtau2}), though this
would require using information from neighboring cells, or we could
avoid the Eddington approximation (eq.~\ref{eq:eddington}) and instead
work directly with the pressure tensor ${\bf K_\nu}$ (which HEROIC
does compute in each cell during each iteration). But we have not
found such refinements necessary in the work we have done so far.

\subsection{Accelerated Lambda Iteration}\label{sec:ALI}

The algorithm used by HEROIC to solve the radiative transfer problem
is Lambda Iteration, but enhanced with a simple form of
acceleration. The basic Lambda Iteration (LI) algorithm is
straightforward and goes as follows. At the end of iteration $n$,
given the current estimate of the source function $S_\nu^n$ as well as
the current estimates of the opacity coefficients, $\kappa_\nu^n$,
$\sigma_\nu^n$, $\epsilon_\nu^n$ (corresponding to the current
estimate of the local temperature) in all the cells, new intensities
for the $(n+1)$th iteration are computed via equation
(\ref{eq:formal}) (for all ray directions at all frequencies in all
cells, see Fig.~\ref{fig:short_char})).  From these intensities, the
new $J_\nu^{n+1}$ is obtained and the Kompaneets equation is solved in
each cell to obtain $a_\nu^{n+1}$. Combined with the new $B_\nu^{n+1}$
corresponding to the current temperature, this enables one to compute
$S_{\nu,\rm LI}^{n+1}$ via equation (\ref{eq:snucompt}), where the
subscript LI is to indicate that this estimate corresponds to LI. If
needed, temperatures are updated at this point, thus completing one
iteration of LI.

In the presence of strong scattering, especially when the scattering
optical depth $\tau$ across a cell is large, LI is very slow to
converge, requiring of order $\tau^2$ iterations.  The solution is to
use Accelerated Lambda Iteration (ALI, see \citealt{hubeny03} for a
short review of the method). Formally, the LI steps described above
may be viewed as a mapping between the new mean intensities
$J_\nu^{n+1}$ in the various cells and the previous source functions
$S_\nu^n$. The mapping is linear and may be formally written via a
$\Lambda$ operator,
\begin{equation}
J_\nu^{n+1} = \Lambda(\{S_\nu^n\}),
\end{equation}
where it must be stressed that $\Lambda$ is not a local relation
within a single cell but couples all cells via the radiative transfer
equation. If the $\Lambda$ matrix can be inverted, then one could
achieve very rapid convergence to the solution.  However, in practice,
the $\Lambda$ matrix is too large and difficult to invert directly, so
an approximate operator $\Lambda^*$ is used instead.

The simplest approximation, the one we use, is to consider only the
diagonal components of the $\Lambda$ operator, motivated by the fact
that the diagonal terms usually dominate the system.  These terms
represent the contribution of $S_\nu^n$ in a given cell to
$J_\nu^{n+1}$ in the same cell, the ``self-illumination contribution''
to the source term.  Thus we rewrite the $\Lambda$ mapping as follows
\begin{equation}\label{eq:lambda}
J_\nu^{n+1} = \lambda_\nu^n S_\nu^n + \Lambda'({S_\nu^{n}}),
\end{equation}
where now $\lambda_\nu^n$ is a number (not a matrix) associated with a
single cell and frequency, $J_\nu^{n+1}$ and $S_\nu^n$ correspond to
the same cell and frequency, and the final $\Lambda'$ term represents
the remaining non-local part of the $\Lambda$ operator that couples
information from other cells. Using the above relation, the updated
$S_{\nu,\rm LI}^{n+1}$ from LI can be written as
\begin{eqnarray}\label{eq:snuli}
S_{\nu,\rm LI}^{n+1} &=& \epsilon_\nu^{n+1} B_\nu^{n+1} + (1+a_\nu^{n+1}) J_\nu^{n+1} \nonumber \\
&=& \epsilon_\nu^{n+1} B_\nu^{n+1} +(1+a_\nu^{n+1}) \lambda_\nu^n S_\nu^n \nonumber \\
&~& \qquad \qquad +(1+a_\nu^{n+1}) \Lambda'(\{S_\nu^{n}\}).
\end{eqnarray}
In the ALI scheme, $S_\nu^n$ in the middle term on the right hand side
is replaced with $S_\nu^{n+1}$ and this term is brought over to the
left hand side. Solving the resulting equation for $S_\nu^{n+1}$ then
gives the ALI estimate for the new source function:
\begin{equation}\label{eq:snuali}
S_{\nu,\rm ALI}^{n+1} = S_\nu^n + \frac{(S_{\nu,\rm LI}^{n+1} -
  S_\nu^n)}{[\, 1 - (1+a_\nu^{n+1}) \lambda_\nu^n \,]}.
\end{equation}
The factor in the denominator in the last term is generally smaller
than unity, which means that the shift in the source function from one
iteration to the next is larger (sometimes much larger) than with
basic LI. This results in accelerated convergence\footnote{For greater
  stability, we usually replace $(1+a_\nu^{n+1})$ by
  $(1-\epsilon_\nu)$ in equation (\ref{eq:snuali}), thereby reverting
  to ALI without Comptonization (see paper 1). Although this slows
  down the rate of convergence, it makes the code more robust.}. Note
that all the quantities except $\lambda_\nu^n$ on the right hand side
of equation (\ref{eq:snuali}) are available at the end of each
iteration of LI, and $\lambda_\nu^n$ itself can be easily computed as
follows.

From equation (\ref{eq:lambda}) it is seen that, for each cell and
frequency, $\lambda_\nu^n = \partial J_\nu^{n+1} / \partial S_\nu^n$
while keeping all the other source function terms constant (i.e.,
keeping the $\Lambda'$ term constant). Therefore, in parallel with the
regular LI calculation, we carry out a second intensity calculation
where, for each ray in equation (\ref{eq:formal}), we set
$I_\nu(\tau_{\nu,1}) = S_\nu(\tau_{\nu,1}) = 0$ and
$S_\nu(\tau_{\nu,2}) = 1$. We recompute the Taylor series
(\ref{eq:snuseries}) for these values of the two source functions,
using the appropriately modified $S''_\nu$. The resulting
$J_\nu^{n+1}$ directly gives $\lambda_\nu^n$.

As discussed earlier, the particular ALI approach described here
focuses just on the self-illumination term and thereby avoids
inverting the full $\Lambda$ matrix. This is a very good approximation
for the spatial part of the problem because ALI is needed most when
the scattering optical depth across a single cell is large, and it is
precisely in this limit that the exponentials in (\ref{eq:formal})
ensure that the intensity is dominated by the local source function.
On the other hand, the decomposition in equation (\ref{eq:lambda})
also isolates frequencies from one another and it is not a good
approximation to assume that the self-illumination term dominates in
frequency space, where the primary effect of Comptonization is to move
radiation from one frequency to another.

We find that HEROIC with the simple version of ALI described here
requires more iterations to converge when tackling a problem involving
Comptonization than for an identical problem with only Thomson
scattering. On the other hand, ALI still gives much faster convergence
than simple LI.  As a final comment, the coupling of frequencies in
the $\Lambda$ matrix could be incorporated into equation
(\ref{eq:lambda}) by making $\lambda_\nu^n$ itself a matrix that
couples neighboring frequencies over a stencil size of order the
frequency dispersion of the Kompaneets operator.  We have not found it
necessary to experiment with such refinements. In rare circumstances,
usually when the Compton $y$-parameter across a cell is very large,
ALI can be unstable.  In these cases, it necessary to switch back to
ordinary LI.

\subsection{Solving for the Temperatures}\label{sec:temperature_solve}

The methods described so far are sufficient to obtain a solution to
the radiation problem, provided the temperature, or equivalently
$B_\nu$, is given in each cell. This is the case for the test problems
described in \S\ref{sec:NumericalTests}. However, in the applications
discussed in \S\ref{sec:disc}, and for many future applications of
HEROIC, we will not know in advance the temperatures in individual
cells but will need to solve for them.

In a typical accretion disc problem, one will obtain for each cell,
say from a GRMHD simulation, the density $\rho$, the four-velocity
$u^\mu$, the temperature $T'$ (the prime here is to indicate that this
is the temperature as determined by the simulation, which we will
improve as part of solving the radiative transfer problem), and the
heating rate per unit volume $Q^+$ (${\rm erg\,cm^{-3}s^{-1}}$). This
last quantity is the rate at which thermal energy is added to the gas
by viscous dissipation. For now, let us ignore energy advection and
assume that $Q^+$ is equal to $Q^-$, the rate at which energy is
transferred from gas to radiation per unit volume:
\begin{equation}\label{eq:qminus1}
Q^- = Q^+.
\end{equation}
This relation can be applied in each spatial cell in the grid, and
this set of equations can be used to solve for the temperature.
Details are given below.

There are two ways of estimating the cooling rate $Q^-$ from the
radiative transfer solution.  First, we can take a microscopic
approach in which we sum up all the intensity added to the radiation
field through thermal emission or Compton scattering, and subtract
from this all the intensity removed by absorption and scattering. The
result, suitably integrated over frequency and angles, is obviously
the net cooling rate of the gas. This gives
\begin{equation}\label{eq:qminus2}
Q^- = 4\pi \int_\nu\left[\kappa_\nu(B_\nu - J_\nu) + \sigma_\nu
  (A_\nu - 1) J_\nu \right] d\nu,
\end{equation}
where the factor of $4\pi$ is to go from ``per steradian'' to the
``whole sphere''. All the quantities on the right are known at each
iteration of the radiative transfer algorithm and therefore it is
possible to estimate $Q^-$ for every cell.

The second estimate of $Q^-$ comes from the spatial divergence of the
radiation flux ${\bf F}$, where the latter can be written in terms of
${\bf H}_\nu$ (eq.~\ref{eq:hnu}) by
\begin{equation}
{\bf F} = 4\pi c\int_\nu {\bf H}_\nu d\nu.
\end{equation}
The energy conservation equation of the radiation field then becomes
\begin{equation}\label{eq:qminus3}
Q^- = \nabla_r \cdot {\bf F},
\end{equation}
where the subscript $r$ on $\nabla_r$ is to indicate that the
divergence is computed in spatial coordinates and not optical depth as
in \S\ref{sec:quadratic}.

The two estimates of $Q^-$ given in equations (\ref{eq:qminus2}) and
(\ref{eq:qminus3}) are guaranteed to be driven towards each other as
the radiation solution converges (subject to accuracy limitations due
to the gridding of the problem). Therefore, we could substitute either
of these estimates in equation (\ref{eq:qminus1}) when solving for the
temperature. The best strategy in our experience is to use equation
(\ref{eq:qminus2}) in optically thin regions and equation
(\ref{eq:qminus3}) in optically thick regions and to transition
smoothly between the two zones (the precise details do not seem to
matter), very much in the spirit of the Lucy-Unsold method
citep{mihalas78}.  A motivatation this choice is that, in the
optically thick regime, the opacities are large and $B_\nu$ and
$J_\nu$ are also very large, but their difference is small. Therefore,
the overall integral in equation (\ref{eq:qminus2}) is much smaller
than the values of the individual terms. Consequently, equation (36)
provides a more accurate method for calculating $Q^{-}$.

Whichever version of $Q^-$ we substitute in equation
(\ref{eq:qminus1}), we need to solve the resulting (coupled)
non-linear set of equations for the temperatures. In optically thin
regions, where we use equation (\ref{eq:qminus2}), a
Newton-Raphson-like approach is sufficient, but optically thick
regions are more difficult. In the latter regions we use the form of
$Q^-$ given in equation (\ref{eq:qminus3}), which ultimately takes the
form of the Laplace equation for the pressure tensor (see
eq.~\ref{eq:nablaK}). We have had some success using acceleration
schemes such as the successive over-relaxation (SOR) method
\citep{press92} to solve this strongly spatially coupled
problem. However, note that the entire problem --- radiation plus
temperature --- is often very involved and requires many iterations
(hundreds to thousands), so one does not gain much by accelerating the
temperature solution. Simple-minded local corrections to the
temperature are frequently sufficient and converge to the correct
temperature solution by the time the Comptonized radiative transfer
problem has converged.

We now discuss the relativistic generalization of the above equations.
The quantities $Q^+$, $Q^-$ and all the terms in the right-hand side
of equation (\ref{eq:qminus2}) are defined in the local comoving fluid
frame. Therefore, when we use this version of $Q^-$, relativity
introduces no modifications. Equation (\ref{eq:qminus3}) is more
troublesome since the divergence operator is best handled in the lab
(Boyer-Lindquist) frame and it is necessary to write the energy
conservation equation of radiation in the latter frame.  The time
component of the energy-momentum conservation law for radiation takes
the form (see \citealt{sadowski13,sadowski15b})
\begin{equation}\label{eq:qminus4}
\left(R_0^\mu\right)_{;\mu} = -G_0,
\end{equation}
where $R^\mu_\nu$ is the stress-energy tensor of the radiation and
$G_\nu$ is the radiation four-force. Both quantities are easily
evaluated in the orthonormal fluid frame, where all our radiation
quantities like $J_\nu$, $B_\nu$, $\kappa_\nu$, $a_\nu$, etc. are
defined. In particular, the fluid-frame radiation four-force has time
component $\widehat{G}_0 = -Q^-$, which is the energy transferred from
the radiation to the gas, and the spatial components of the four-force
are similarly the momentum components transferred from radiation to
gas.  Once $\widehat{R}^\mu_\nu$ and $\widehat{G}_\nu$ are computed in
the fluid frame, one transforms these tensors and vectors to the lab
frame and then substitutes the appropriate components into equation
(\ref{eq:qminus4}). This is the relativistic generalization of
equation (\ref{eq:qminus3}).

Finally, we discuss the issue of energy advection. Equation
(\ref{eq:qminus1}) is valid only for radiatively efficient flows where
the gas immediately radiates whatever heating it experiences.  In the
more general situation, there is an additional entropy advection term
in the energy equation and we have
\begin{equation}\label{eq:qminus5}
Q^- = Q^+ - \rho T \frac{ds}{dt} = Q^+ - \frac{\rho kT}{\bar{m}} \frac{d}{dt}
\left(\ln\frac{T^n}{\rho}\right).
\end{equation}
Here $d/dt$ represents a Lagrangian time derivative following a fluid
element, $\bar{m}$ is the mean mass per particle in the fluid, and $n
= 1/(\Gamma - 1)$ is the polytropic index of the gas. This is a
nonrelativistic version of the equation.

Since $d/dt$ is a Lagrangian time derivative, we can rewrite equation
(\ref{eq:qminus5}) in four-notation as
\begin{equation}\label{eq:qadv}
Q^- = Q^+ - \frac{c}{R_g} \left(\frac{n\rho k}{\bar{m}} u^\mu T_{;\mu} -
\frac{kT}{\bar{m}} u^\mu \rho_{;\mu} \right),
\end{equation}
where the additional factor of $c/R_g$ is to convert from
gravitational units used in typical GRMHD simulations to physical
units; here $R_g = GM/c^2$, where $M$ is the mass of the black hole
(or other central gravitating object). Specializing further to
Boyer-Lindquist coordinates, assuming steady state (no time dependence
of quantities) and axisymmetry (no $\phi$ dependence), and replacing
semicolons by commas since $T$ and $\rho$ are scalars, we obtain:
\begin{equation}
Q^- = Q^+ - \frac{nck\rho}{\bar{m}R_g} \left(u^r T_{,r} + u^\theta
T_{,\theta}\right) + \frac{ckT}{\bar{m}R_g} \left(u^r \rho_{,r} +
u^\theta \rho_{,\theta}\right).
\end{equation}
This is the form of the energy equation (rather than
eq.~\ref{eq:qminus1}) that HEROIC uses when solving for
temperatures. In practice, it makes a difference only in regions where
the gas has a tendency to be advection-dominated, e.g., the plunging
region of an accretion flow inside the ISCO (see \citealt{zhu12}) or
the funnel region where a jet may be present.

\subsection{Long Characteristics and Ray Tracing}\label{sec:raytrace}

The letter H in the name of the code HERO described in Paper 1 stands
for Hybrid and refers to the fact that the code combines several
stages: an initial stage in which the problem is solved using a short
characteristics solver, a second stage in which the solution is
improved using a long characteristics solver, and a final stage in
which ray tracing is done to compute observables for a distant
observer.

In the case of HEROIC, all the discussion so far was related to the
short characteristics method. Unfortunatley, we do not yet have a long
characteristics solver that can handle Comptonization. We do, however,
have the ray tracing code that was already developed for HERO (Paper
1), and the same code can be used even when there is Comptonization.
This is because the short characteristics solver obtains a solution
for the source function $S_\nu$ and the temperature $T$ (which allows
one to calculate opacities). This is all that one needs for
ray-tracing, since the latter involves nothing more than integrating
equation (\ref{eq:formal}) backwards from a distant observer over a
grid of impact parameters and frequencies.

\section{Numerical Tests}\label{sec:NumericalTests}

We have validated our Compton module in HEROIC by applying it to two
test problems that admit analytic or quasi-analytic solutions.
For these test problems, we considered the case of pure scattering
with no absorption. We also carried out other tests to check some
relativistic aspects of the code.

\subsection{Kompaneets Solver}

We begin by benchmarking the Kompaneets solver, which is the workhorse
that HEROIC relies on to handle all Comptonization problems.  Consider
the ``Green's function'' of an initially monoenergetic distribution of
photons (delta-function in $\nu$) as it evolves in energy via Compton
scattering within a closed box of hot thermal electrons.
\begin{figure}
\begin{center}  
\includegraphics[width=0.5\textwidth]{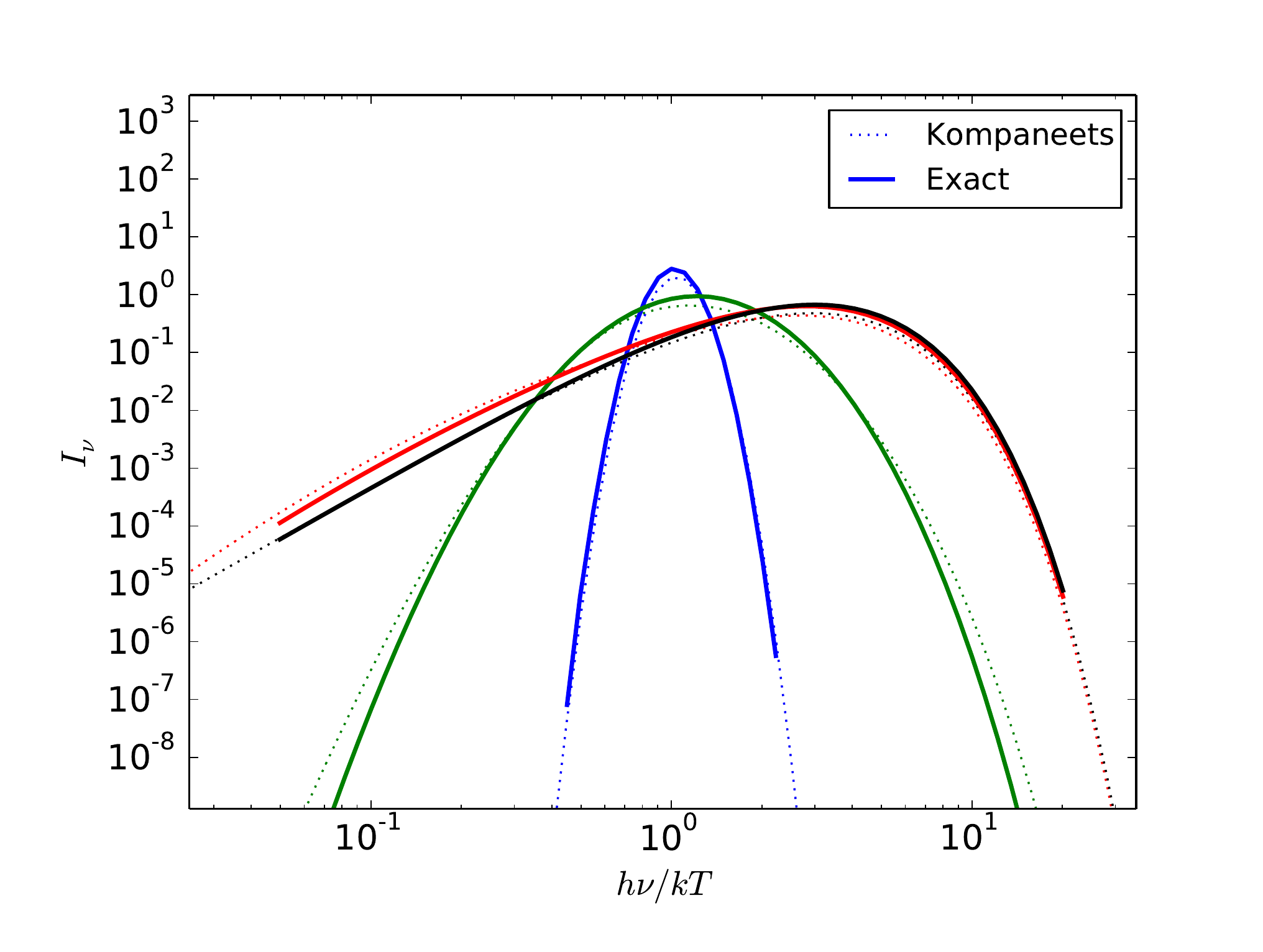}
\caption{Comparison of numerical results obtained with the
  Kompaneets solver employed in HEROIC (\S\ref{sec:KompRay}) with the
  exact analytical solution (Eq.~\ref{eq:Whittaker}), for the spectral
  evolution of an initial delta-function at frequency $x=h\nu/kT=1$.
  The different colors correspond to different values of the Compton
  y-parameter: 0.1 (blue), 1 (green), 10 (red), 100 (black).  The
  slight deviations are largely because the initial distribution in
  the numerical solution is not a perfect delta-function but has a
  finite width equal to the size of a frequency bin.
\label{fig:whittaker}}
\end{center}  
\end{figure}  
If we ignore the nonlinear $n^2$ term in the Kompaneets equation
(\ref{eq:Kompaneets}), the problem is simple enough that an analytical
solution is available in terms of complex integrals of the Whittaker
functions $W_{k,m}(x)$ \citep{becker03}:
\begin{eqnarray}
f_G(x, y) &=&\frac{32}{\pi}e^{-9y/4}x_0^{-2}x^{-2} e^{(x_0-x)/2}\,\times
\nonumber \\ &~& \int \limits_0^\infty e^{-u^2y} \frac{u \,
  \text{sinh}(\pi u)}{(1+4u^2)(9+4u^2)} W_{2,iu}(x_0)W_{2,iu}(x) du
\nonumber \\ &~& + \frac{e^{-x}}{2} +
\frac{e^{-x-2y}}{2}\frac{(2-x)(2-x_0)}{x_0 x},\label{eq:Whittaker}
\end{eqnarray}
where $x=h\nu/kT$ is the dimensionless frequency, $f_G(x,y)$
represents the Green's function spectral response of the system to a
delta source injected at frequency $x_0$ after it has evolved over a
timescale corresponding to a Compton $y=4\theta_e n_{\rm scatt}$.

Figure (\ref{fig:whittaker}) compares numerical solutions from our
Kompaneets solver (\S\ref{sec:KompRay}) to the above analytical
solution for different choices of $y$. In these calculations, we
initialized HEROIC with a photon distribution that is non-zero in a
single frequency bin at $x = x_0 = 1$.  The numerical solutions from
HEROIC agree very well with the analytical solution.

\begin{figure}
\begin{center}  
\includegraphics[width=0.5\textwidth]{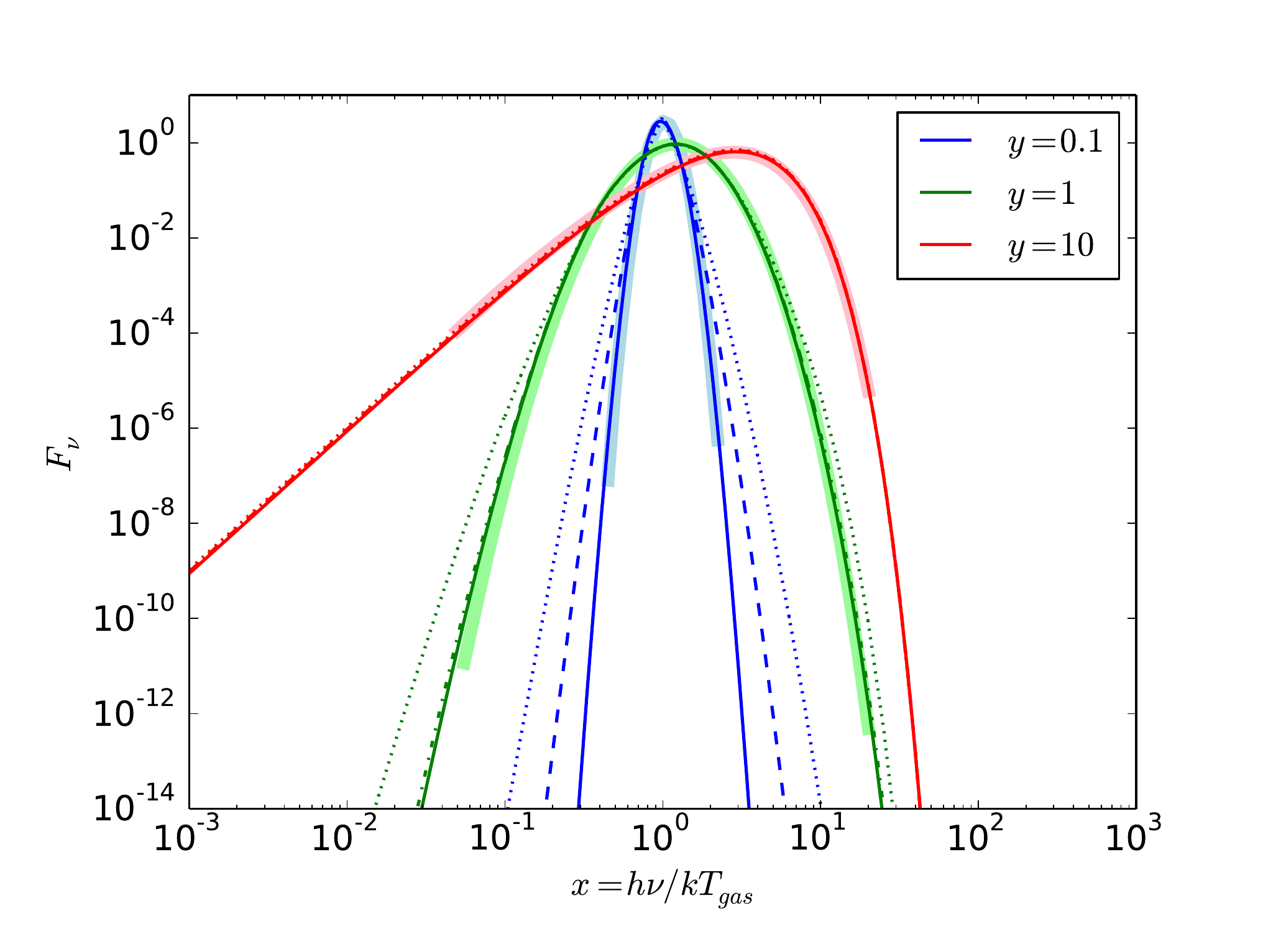}
\caption{Comparison of numerical results from the Kompaneets solver
  with the analytical result for different frequency resolutions.  The
  dotted, dashed and solid lines correspond to 10, 40 and 160 bins per
  frequency decade, respectively.  The thick solid lines show the
  analytical solutions (Eq.~\ref{eq:Whittaker}).
\label{fig:freqRes}}
\end{center}  
\end{figure}

The minor discrepancies seen in Figure \ref{fig:whittaker} are
primarily due to the fact that the initial frequency distribution is
not a perfect delta-function but has a finite width due to the bin
size.  This finite size slightly fattens the numerical Kompaneets
result.  Figure \ref{fig:freqRes} shows the effect of changing the
frequency resolution.  We see that with increasing frequency
resolution the agreement steadily improves.  But even a resolution of
10 frequencies per decade (our default) is sufficiently accurate for
most purposes since the deviations are noticeable only deep in the
wings and that too only for a delta-function (the radiation sources we
deal with generally do not have such narrow spectral distributions).

\subsection{Plane Parallel Comptonizing Atmosphere}\label{sec:planecompt}

A typical Comptonization problem involves both spatial diffusion and
Compton-scattering in frequency. Here we consider a simple problem in
which we have a one-dimensional, plane parallel, hot, scattering
atmosphere with a finite optical depth $2\tau_0$ through it. We assume
steady injection of soft photons with a blackbody spectrum at the
mid-plane. The injected photons random-walk away from the mid-plane,
and their energies become modified with each scattering. We are
interested in the steady state spectrum of the escaping radiation from
the two surfaces.

Using the Kompaneets solver that was tested in the previous
sub-section, we first calculate the correct solution to this model
problem. For this, we need to compute the photon escape time
distribution, where by time we mean the number of scatterings
experienced by a photon before it escapes.  A convolution of the
photon escape time distribution with the Kompaneets derived spectrum
for each escape time then yields the emergent spectrum for the
problem.

\begin{figure}
\begin{center}  
\includegraphics[width=0.5\textwidth]{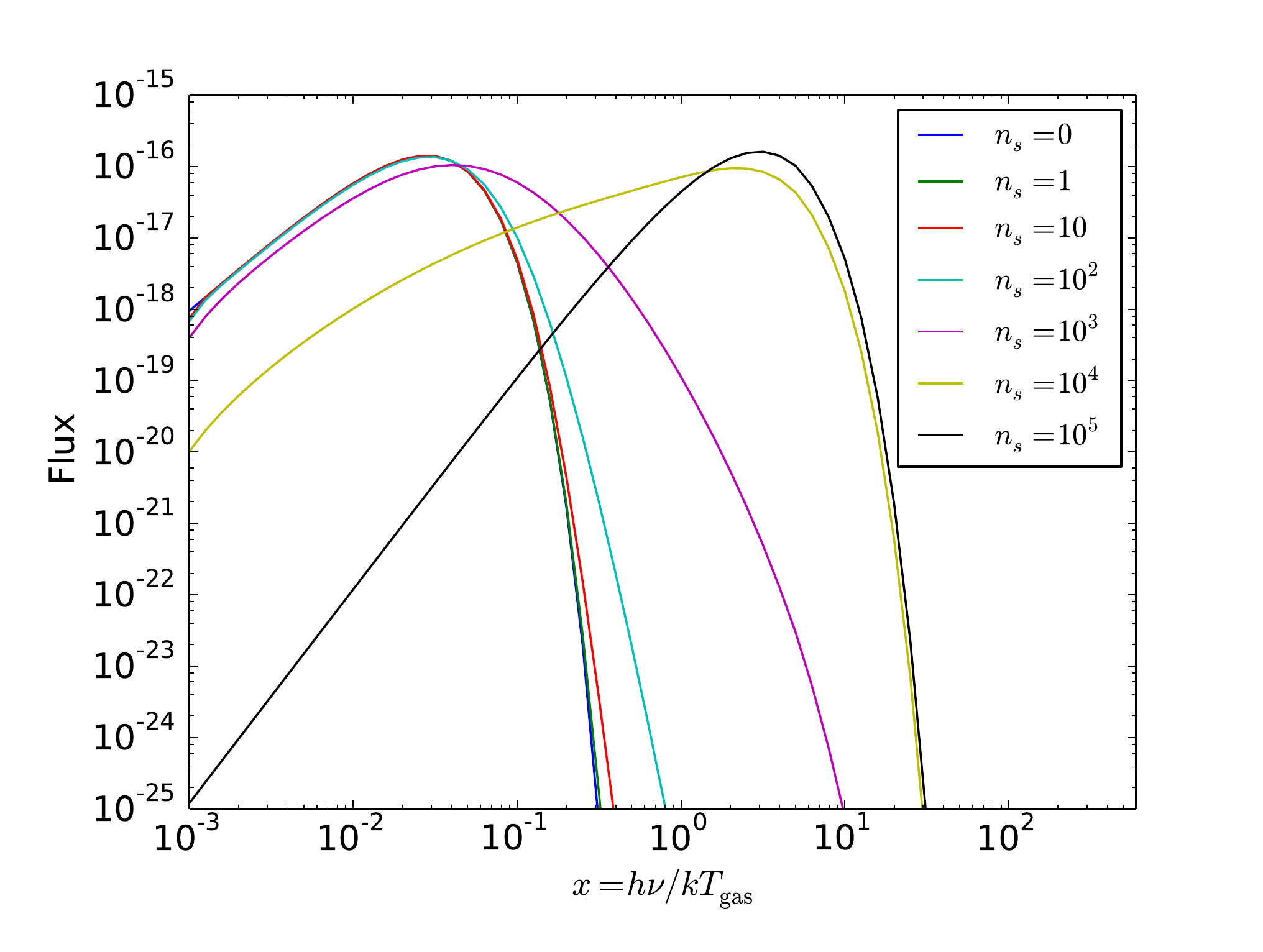}
\caption{Sample Compton kernels computed with the Kompaneets solver.
  The problem considered is the Comptonization of an initial $T=10^4
  K$ radiation field that is upscattered by hot gas with $T=10^6 K$.
  The different curves show Comptonized spectra after various numbers
  $n_s$ of scattering events.
\label{fig:1Dspectra}}
\end{center}  
\end{figure}

Figure \ref{fig:1Dspectra} shows an example of the spectral evolution
of a $T_{\rm rad}=10^4 K$ initial radiation field as it scatters in a
thermal gas with $T=10^6 K$.  This set of spectra constitutes the
``Compton kernel'' which we will use for calculating the emergent
spectrum from a Comptonizing atmosphere.

To calculate the escape time distribution for a 1D plane-parallel
atmosphere, we solve the diffusion equation, i.e.
\begin{equation}
\frac{\partial n}{\partial t} = \frac{1}{3} \frac{\partial^2
  n}{\partial \tau^2},
\end{equation}
where the factor of $1/3$ is because we are considering diffusion of
photons in three dimensions\footnote{This coefficient is also
  compatible with the two-stream approximation for radiative transfer
  \citep{rybicki79}}, $\tau$ is the perpendicular optical depth in the
atmosphere, and $t$ is measured in units of the number of scattering
times. We initialize the system with a spatial delta-function
distribution of photons at the mid-plane.  We then allow the system to
evolve and calculate the escaping photon number flux at the surface as
a function of time. This flux, normalized by the number of initial
photons, directly gives the escape time probability distribution.

\begin{figure}
\begin{center}  
\includegraphics[width=0.5\textwidth]{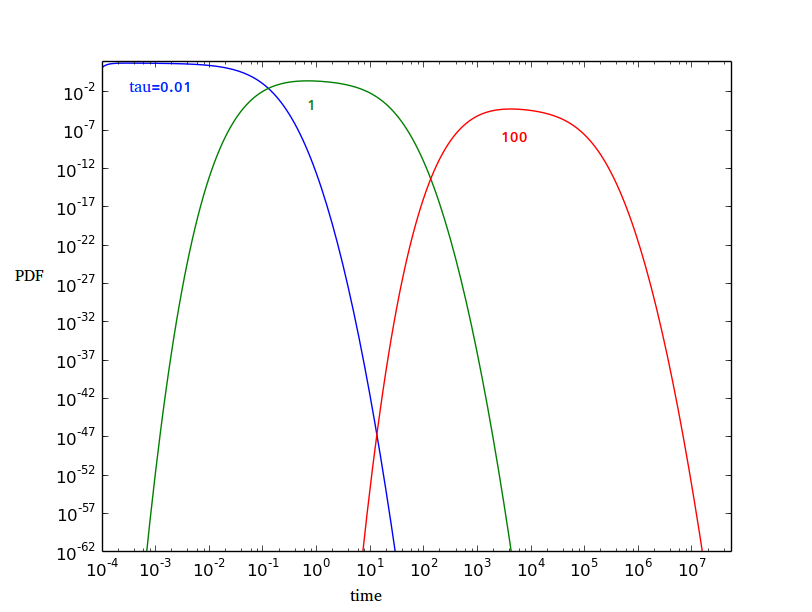}
\caption{The escape time probability distribution function for various
  choices of plane parallel slab thickness.  The escape time is
  measured in units of the characteristic scattering time.  Note that
  the diffusion time scales with optical depth as $t\sim\tau$ in the
  optically thin limit and as $t\sim\tau^2$ in the optically thick
  limit.
\label{fig:1Dkernel}}
\end{center}  
\end{figure}

\begin{figure}
\begin{center}  
\includegraphics[width=0.5\textwidth]{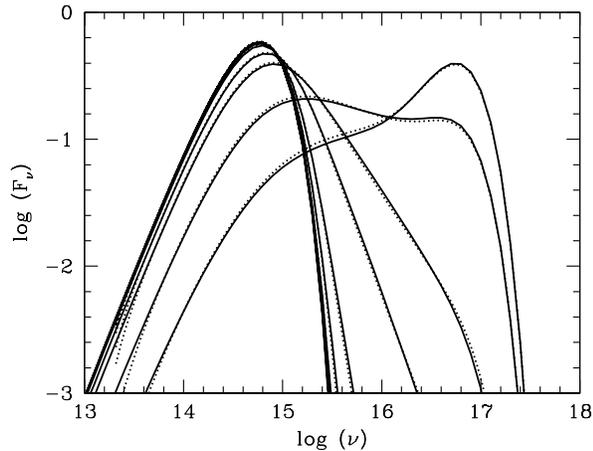}
\vspace{-1.8 in}
\caption{Spectra of the emergent radiation in the 1D plane parallel
  Compton scattering problem.  Blackbody radiation with a temperature
  of $10^4$\,K is injected at the midplane of the slab and the
  scattering electrons are assumed to have a temperature of
  $10^6$\,K. From left to right as measured by the positions of the
  peak, the scattering optical depths to the midplane of the slab are:
  $\tau=0$, 1, 2, 3, 6, 10, 20, 30, 60, 100, respectively.  Solid
  lines represent results obtained with (1D) HEROIC using two rays
  (which corresponds to the 2-stream approximation) and dotted lines
  represent exact results computed as described in the text.
\label{fig:planeparallelspectra}}
\end{center}  
\end{figure}

Instead of taking the approach of \citet{sunyaevtitarchuk80}, who
tackle the diffusion problem analytically via series expansions, we
opt for a numerical solution.  We consider the upper half of the slab,
and apply a reflecting boundary condition at the midplane because of
symmetry.  At the surface, for simplicity, we use the boundary
condition appropriate to the two-stream approximation:
\begin{equation}\label{eq:bc}
\left. \frac{\partial n}{\partial \tau} \right|_{\rm surf} + \sqrt{3}
\left. n \right|_{\rm surf} = 0.
\end{equation}
Figure \ref{fig:1Dkernel} shows the calculated distributions of escape
times for a few values of $\tau$.  Convolving these with the Compton
kernel shown in Figure \ref{fig:1Dspectra} then gives the energy
spectra of the escaping photons as a function of frequency. The
results are shown by the dotted lines in Figure
\ref{fig:planeparallelspectra} for model atmospheres of various
optical depths. The temperature of the scattering atmosphere is taken
to be $10^6$\,K, and the photons injected at the mid-plane have a
blackbody distribution with a temperature of $10^4$\,K.

Having computed the correct solutions, we solved the same problem
using HEROIC. We used a logarithmically spaced spatial grid of 101
points, with 20 points for each decade of optical depth. In order to
be consistent with the boundary condition (\ref{eq:bc}) we used only
two angles, and we used 61 frequencies distributed uniformly in
$\log\nu$ from $\nu=10^{12}$\,Hz to $10^{18}$\,Hz. We treated the
midplane as a reflecting boundary with an additional steady source of
blackbody radiation with temperature $10^4 K$. At the outer surface of
the atmosphere, we assumed that there is no ingoing radiation. Since
the temperature of the gas in the atmosphere is given ($T = 10^6$\,K),
there is no need for the temperature solution methods described in
\S\ref{sec:temperature_solve}.

The resulting spectra obtained with HEROIC are shown by the solid
lines in Figure~\ref{fig:planeparallelspectra}. A comparison of these
with the true solutions (dotted lines) shows that the agreement is
quite good. Note that this is a comprehensive test of Comptonization
at nonrelativistic temperatures since it includes all the elements
described in \S\S\ref{sec:KompRay}, \ref{sec:quadratic} and
\ref{sec:ALI}. It is, however, a one-dimensional problem, whereas
HEROIC was developed principally for multi-dimensional problems.

\subsection{Spherical Scattering Atmosphere}\label{sec:sphericalcompt}

Here we consider a homogeneous spherical Comptonizing atmosphere where
photons are injected at the center and diffuse outwards.  This problem
is identical to the plane parallel previously discussed, except that
it now occurs in spherical geometry.  The corresponding diffusion
equation is \citep{sunyaevtitarchuk80}:
\begin{equation}
\frac{\partial n}{\partial t} = \frac{1}{3} \frac{1}{\tau^2}
\frac{\partial}{\partial \tau}\left( \tau^2 \frac{\partial n}{\partial
  \tau}\right).\label{eq:3DdiffusionEq}
\end{equation}
We inject photons at the center (in practice at a radius $r_{\rm
  min}=10^{-4} r_{\rm max}$) and solve the diffusion problem
numerically for different radial optical depths $\tau_0$. This gives
us the escape time probability distributions.  The central boundary
condition is still set as reflection at $r=r_{\rm min}$ (to simulate
the symmetry at the origin), while the outer boundary condition
becomes modified to (see \citealt{sunyaevtitarchuk80} -- Appendix A):
\begin{equation}
\left. \frac{\partial n}{\partial\tau} \right|_{\rm surf} +
\frac{3}{2} \left. n \right|_{\rm surf} = 0.
\end{equation}

\begin{figure}
\begin{center}  
\includegraphics[width=0.5\textwidth]{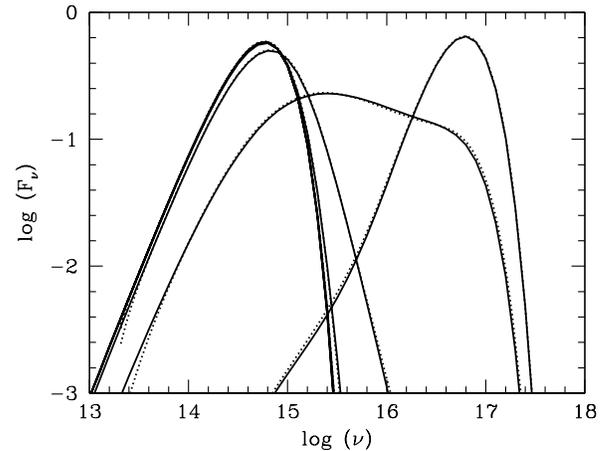}
\vspace{-1.8 in}
\caption{Same as figure \ref{fig:planeparallelspectra}, but for a 3D
  spherical diffusion problem.  The injected radiation temperature and
  the scattering electron temperature are $10^4$\,K and $10^6$\,K
  respectively. From left to right in the position of the peak, the
  radial scattering optical depth of the spherical atmosphere is $\tau
  = 0$, 1, 3, 10, 30, 100, 300, respectively.  Solid lines represent
  results obtained with 2D axisymmetric HEROIC using 80 angles and
  dotted lines represent exact results.
\label{fig:3Dspectra}}
\end{center}  
\end{figure}

As before, we calculate the escape time probability distribution by
solving the diffusion equation numerically and we then convolve this
with the Compton kernel to compute the energy spectrum of escaping
photons. The dotted lines in Figure \ref{fig:3Dspectra} show the
results. Note that the spectra are qualitatively quite similar to
those for the 1D problem (Fig.~\ref{fig:planeparallelspectra}), but
the 3D spherical diffusion problem has on average shorter escape times
compared to an equivalent 1D plane parallel problem with the same
scattering depth.  This is a simple consequence of the geometry -- the
mean distance to the surface is shorter in the 3D case.  The shorter
escape times in the spherical case translate to less strongly
Comptonized spectra, as can be seen by comparing the dotted lines in
Figures \ref{fig:3Dspectra} and \ref{fig:planeparallelspectra} for the
same optical depth.

We solved the above 3D spherical problem independently using HEROIC.
We assumed axisymmetry and solved for the radiation field in the 2D
poloidal plane in spherical coordinates $r$, $\theta$ (there is no GR
in this problem). The numerical grid consisted of 61 cells in radius
distributed uniformly in $\log r$ from $r_{\rm min}=3$ to $r_{\rm
  max}=300$, and 31 points distributed uniformly in $\theta$. The
radiation field was solved on 80 angles distributed uniformly in
direction (see Paper 1 for details) and 61 points in frequency exactly
as in the 1D problem.  To be consistent with the choice of a constant
diffusion coefficient in Equation~(\ref{eq:3DdiffusionEq}), the
scattering sphere was taken to have a constant scattering opacity. The
solid lines in Figure~\ref{fig:3Dspectra} show the results. We see
that the agreement with the true spectra (dotted lines) is as good as
in the plane parallel case. This test shows that Comptonization in
HEROIC works correctly in multiple dimensions.

It should be noted that the short characteristics method used for
these calculations has a serious ray defect in spherical coordinates,
for which Paper 1 developed an approximate fix.  Although this fix
mitigates the error by a large factor, it is imperfect. In particular,
the luminosity as a function of radius in a spherical problem is not
constant (see Fig.~17 in Paper 1 and also
Fig.~\ref{fig:gravitationalredshift} below). In the context of the
present Comptonization test problem, the ray defect results in the
photon luminosity (photons per second) escaping at $r = r_{\rm max}$
being somewhat less than the luminosity injected at $r = r_{\rm
  min}$. For the spectra shown by solid lines in Figure
\ref{fig:3Dspectra}, we have normalized the escaping spectra by a
correction factor so that the net photon luminosity is the same as in
the corresponding model solutions (dotted lines). The energy spectra
were not adjusted in any other way, so the very good agreement in the
shapes of the spectra is a strong test of the algorithm used in
HEROIC.

\subsection{Gravitational Redshift}

HEROIC is designed to work in relativistic spacetimes, and Paper 1
described a number of tests to verify that that HERO correctly handles
ray propagation in vacuum. The tests included light-bending,
gravitational redshift and Doppler shift. In this and the following
subsection, we consider non-vacuum tests.

\begin{figure}
\begin{center}  
\includegraphics[width=0.5\textwidth]{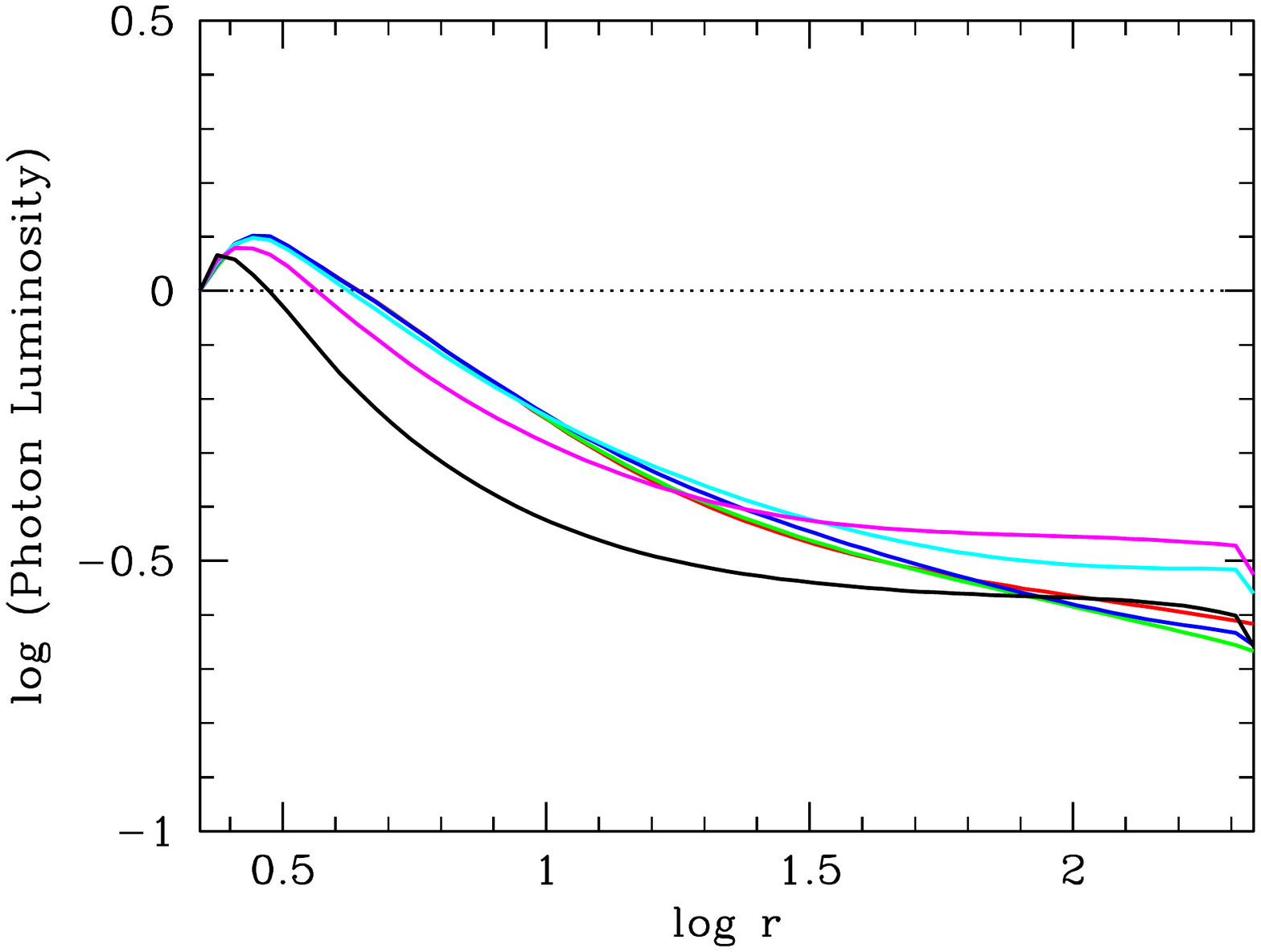}\vspace{-1.8 in}
\includegraphics[width=0.5\textwidth]{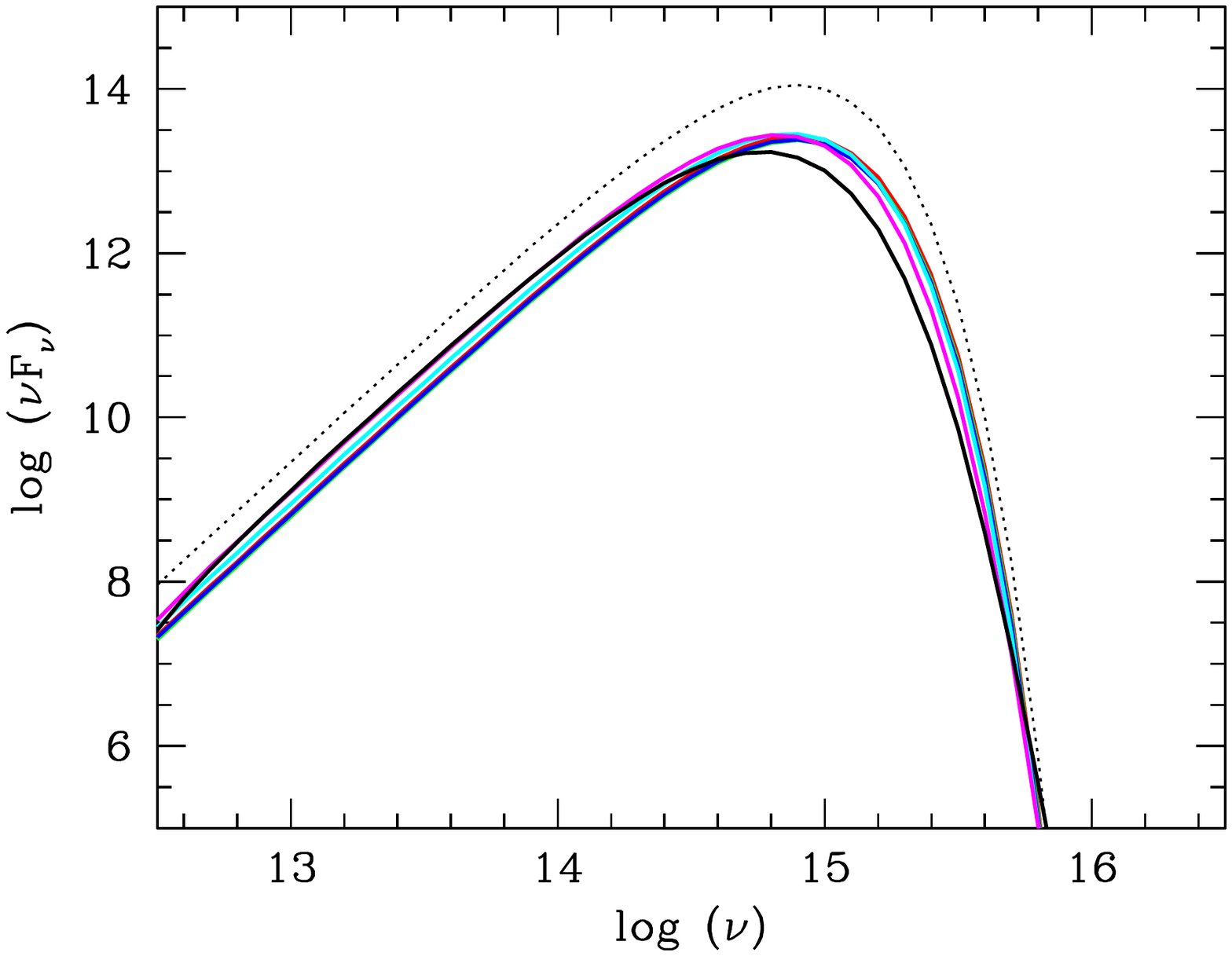}\vspace{-1.5 in}
\caption{Top: Radial photon luminosity (photons/sec) normalized to the
  injected luminosity at the inner edge ($r_{\rm min} = 2.2$), as a
  function of radius $r$ (in units of $GM/c^2$), for different choices
  of the radial scattering optical depth: $\tau = 1$ (red line), 3
  (green), 10 (blue), 30 (cyan), 100 (magenta), 300 (black). Bottom:
  Corresponding spectra of the escaping radiation at the outer edge
  ($r_{\rm max} = 100\, r_{\rm min}$).}
\label{fig:gravitationalredshift}
\end{center}  
\end{figure}  

Figure \ref{fig:gravitationalredshift} shows results corresponding to
radiation propagating in a homogeneous scattering atmosphere in a
Schwarzschild spacetime. For this test, we assume that a spherical
surface at radius $r_{\rm min} = 2.2GM/c^2$ (where $M$ is the mass)
radiates as a blackbody with ``temperature-at-infinity'' of $10^4$\,K,
i.e., with a local temperature, $T_{\rm surface} = 10^4{\rm
  K}/[1-(2GM/c^2r_{\rm min})]^{1/2} = 3.32\times10^4$\,K, and that the
radiation propagates through a uniform spherical scattering atmosphere
(no absorptive opacity) extending from $r = r_{\rm min}$ up to $r =
r_{\rm max} = 100r_{\rm min}$. We vary the radial optical depth
$\tau_0$ of the atmosphere over a wide range of values up to a maximum
of $\tau_0=300$. The atmosphere is at rest in the lab frame and it is
cold, so there is no Comptonization. (We are not aware of any good
test problems involving Comptonization in relativistic spacetimes.)

We use HEROIC to solve for the radiation field on a grid of 61
logarithmically spaced points in $r$ and 31 uniformly spaced points in
$\theta$, assuming axisymmetry. We use 80 angles and 61 frequencies
distributed uniformly in $\log\nu$ from $\nu=10^{12}$\, Hz to
$\nu=10^{18}$\, Hz.  The upper panel in Figure
\ref{fig:gravitationalredshift} shows the photon luminosity as a
function of radius in the solutions obtained with HEROIC for different
choices of $\tau$. Notice that the luminosity is not constant with
radius. This is because of the ray defect discussed in the previous
subsection. A comparison with Figure 17 in Paper 1 shows similar
features, viz., a drop in the luminosity over the first factor of
several in radius, after which the luminosity remains constant. There
are two differences in the present test.  First, we now have a
scattering atmosphere, not vacuum, so this test verifies that the
approximate correction for the ray defect that was described in
\cite{zhu15} works also in the presence of scattering. Second, the
inner radius here is close to the horizon and well inside the photon
orbit. This means that the majority of rays in the innermost region
are pulled in towards the horizon and only a few rays propagate to
larger radii. HEROIC is not handicapped by such extreme ray
deflections.

The second aspect of this test is to check the spectrum of the
radiation that emerges from the outer edge of the atmosphere and to
verify that it does correspond to a blackbody at $10^4$\,K. The lower
panel in Figure \ref{fig:gravitationalredshift} shows the results.  As
already mentioned, the emerging flux is lower than expected, causing
all the calculated spectra (solid lines) to lie below the
theoretically expected spectrum (dotted line). However, the shapes are
correct. The only exception is the model with the largest $\tau = 300$
where the spectrum (black line) is a little cooler than it should
be. In this last case, the radiation effectively scatters
$\sim\tau^2\sim10^5$ times before emerging at the surface. This test
shows that the code is able to preserve spectra while propagating
through optically thick media in a background with variable
gravitational redshift.

\begin{figure}
\begin{center}
\includegraphics[width=0.55\textwidth]{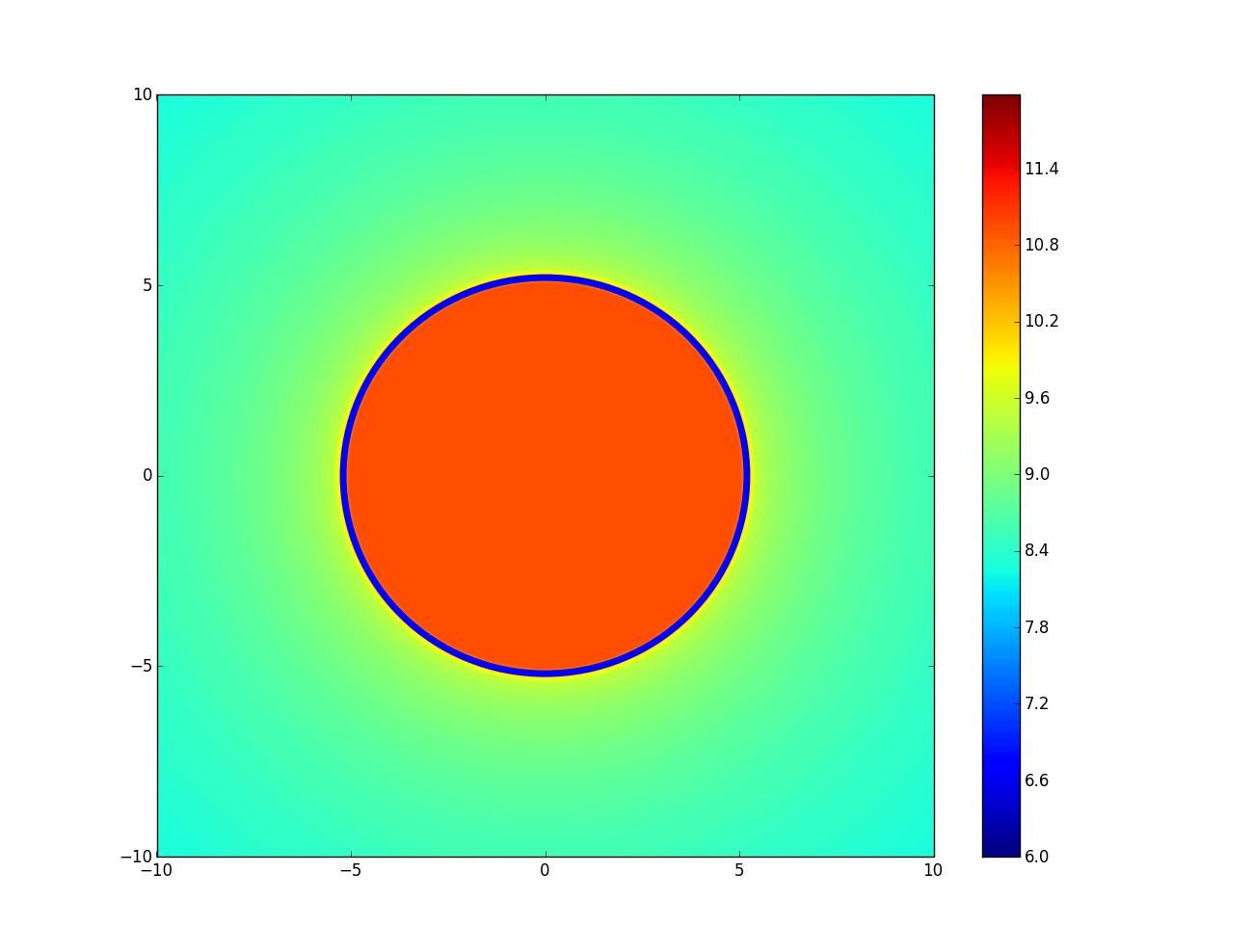}
\caption{Ray-traced image of the model corresponding to $\tau=1$ in
  Figure \ref{fig:gravitationalredshift}. The solid red region shows
  radiation escaping directly from the inner surface of the grid
  ($r=r_{\rm min}$), attenuated by $\exp(-\tau)$.  The
  yellow-green-cyan regions show scattered radiation from the
  scattering atmosphere. Since $r_{\rm min}$ lies inside the photon
  orbit, the apparent size of the inner radiating surface is equal to
  that of the photon orbit ($r_{\rm photon,app} = \sqrt{27}$), shown
  by the blue circle.}
\label{fig:image_tau1}
\end{center}
\end{figure}

Figure \ref{fig:image_tau1} shows the image of this scattering
atmosphere as seen by a distant observer for the model with
$\tau=1$. The bright inner surface, attenuated by a factor of
$\exp(-\tau)$ by scattering, is seen at the center, surrounded by faint
emission from the extended scattering atmosphere. The apparent size of
the inner surface agrees well with the theoretically expected radius
of $\sqrt{27}GM/c^2$, the apparent size of the photon orbit (blue
circle).

\subsection{Radiation Trapping, Bulk Comptonization}

\begin{figure}
\begin{center}
\includegraphics[width=0.5\textwidth]{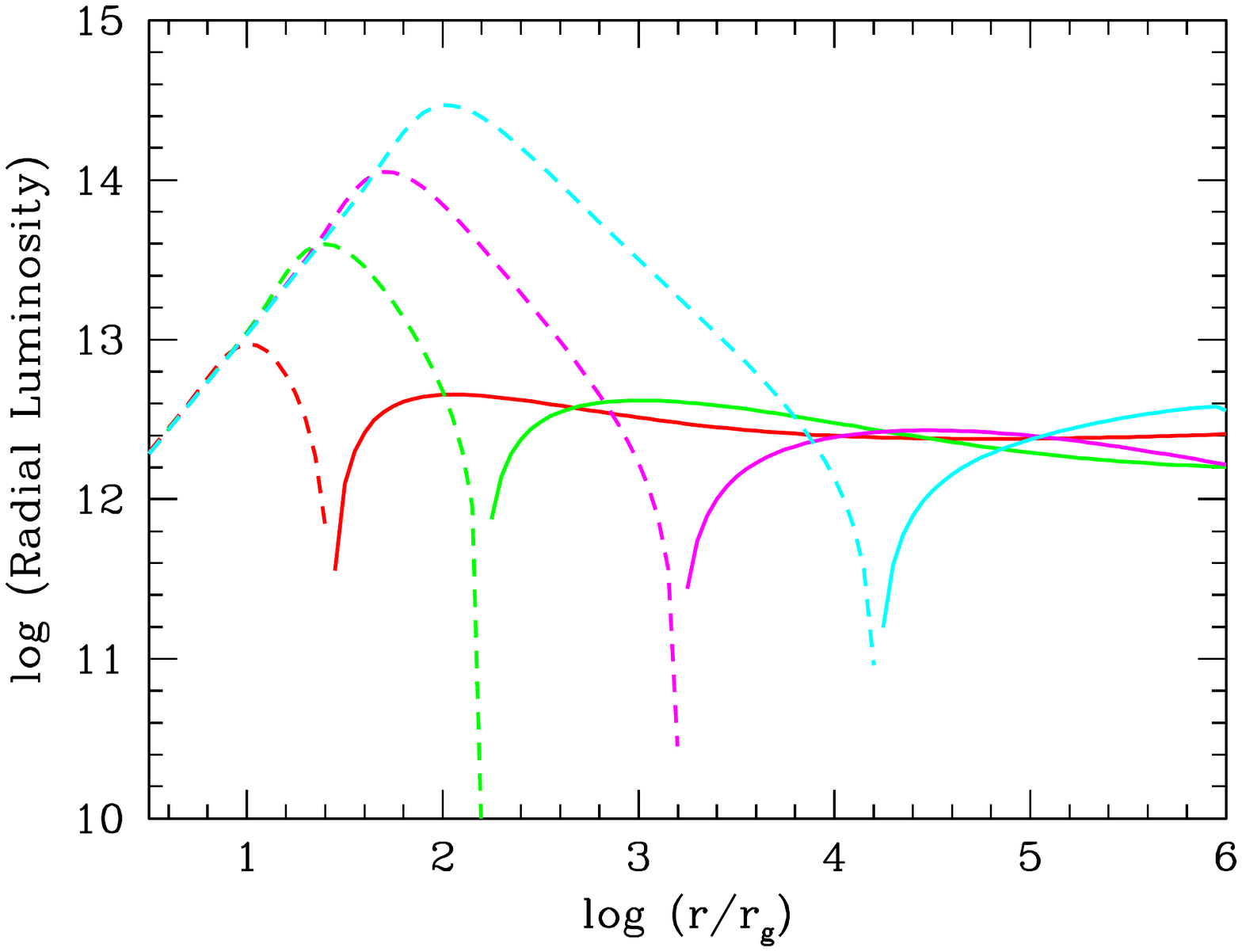}\vspace{-1.8 in}
\includegraphics[width=0.5\textwidth]{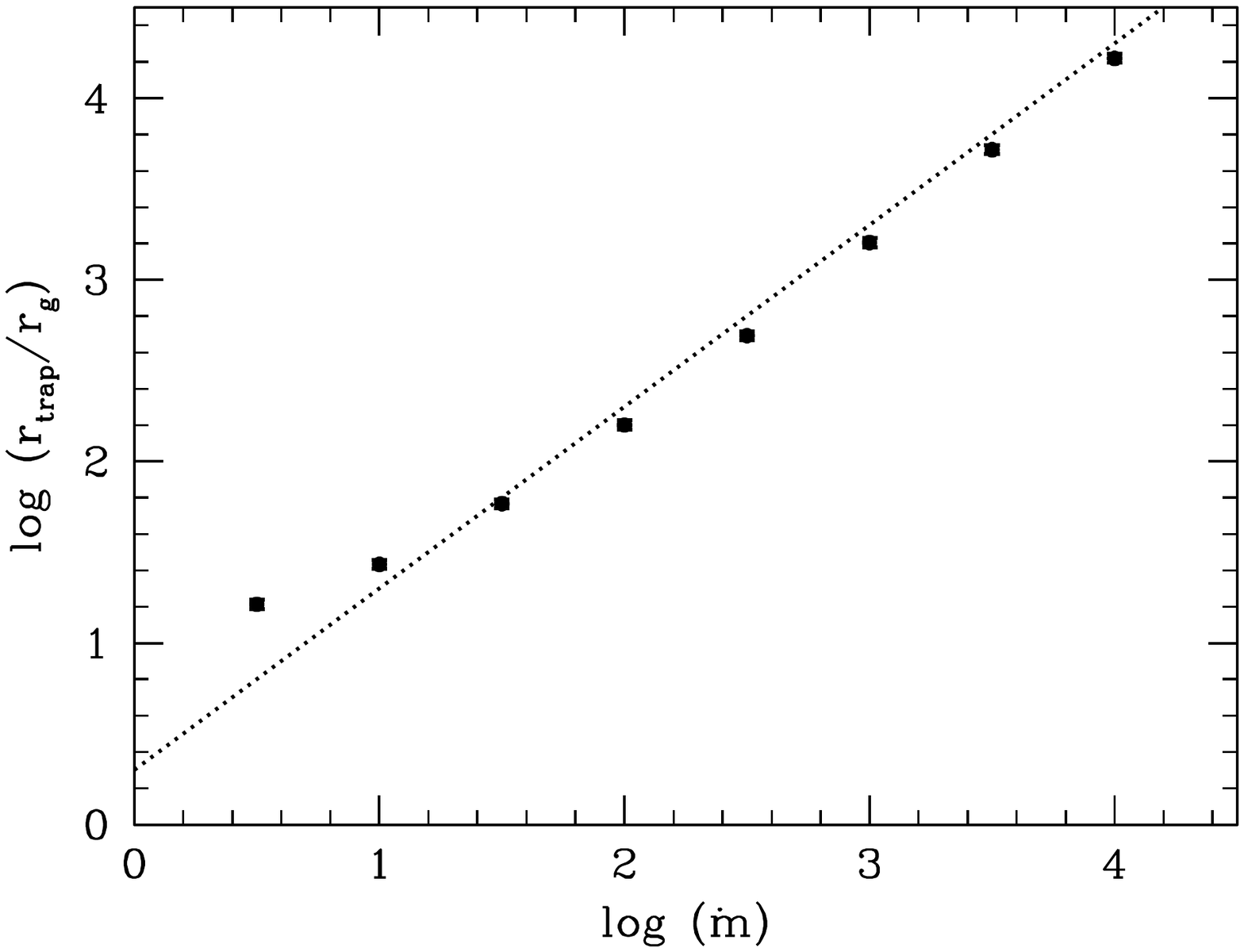}\vspace{-1.5 in}
\caption{Top: Radial photon luminosity (photons/sec) in arbitrary
  units vs radius in units of $r_g$ for spherical Bondi accretion as
  in \citet{turolla02}. Four models are shown, corresponding to
  Eddington-scaled mass accretion rates of $\dot m = 10$ (red), $10^2$
  (green), $10^3$ (magenta), $10^4$ (cyan). Dashed line segments
  correspond to negative luminosities, where radiation is trapped by
  the accreting gas and is dragged into the center, and solid line
  segments correspond to positive luminosities, where radiation flows
  outward. The transition between the two corresponds to the trapping
  radius $r_{\rm trap}$.  Bottom: Variation of $r_{\rm trap}$ with
  $\dot m$. Note the linear dependence once $\dot m \gg 1$.}
\label{fig:rtrap}
\end{center}
\end{figure}

Apart from gravitational redshift and ray deflections, Doppler effects
play an important role in radiative transfer.  The effect we are
interested in here is the advection of radiation by an optically thick
medium. Recall that, in HEROIC, the radiation field is described in
the comoving frame of the fluid, while the radiative transfer
computation is done entirely in the lab frame.  If the fluid moves
with respect to the lab frame, the dragging or advection of radiation
by the moving fluid must ultimately result from Doppler modifications
of the radiation intensity and frequency in the lab frame. We test
this aspect of HEROIC.

The problem we consider is spherical accretion in a Schwarzschild
background. We consider super-Eddington accretion rates so that
radiation is trapped within a certain trapping radius $r_{\rm trap}$
and is dragged to the center. Any radiation outside the trapping
radius is able to escape to infinity. We model this problem as closely
as possible using the setup described in \cite{turolla02}. The
accreting gas has both (grey) absorption and scattering, whose
magnitudes are tuned such that for all models the ``absorption
radius'' $r_a=5M$ and the ``crossing radius'' $r_c=3.6M$, as in
\citet[note that their unit of length is $2M$, not $M$]{turolla02}.
For the opacity index $n$ we choose the middle of the three values
they considered: $n=4$.

Figure \ref{fig:rtrap} shows results obtained with HEROIC for various
choices of the Eddington-scaled mass accretion rate $\dot{m}$.  These
models were computed on a 2D grid with 114 cells distributed
logarithmically in radius, going from $r=10^{0.35}M$ to $r=10^6M$, and
21 cells distributed uniformly in $\theta$; the models assume
axisymmetry (as in all the previous spherical tests), and use 80
angles and 61 frequencies. The upper panel shows the radial luminosity
profiles as measured in the lab frame for $\dot{m} = 10, ~10^2, ~10^3,
~10^4$. Dashed segments correspond to negative luminosities, i.e., the
radiation here is trapped in the accreting gas and dragged inward. The
radius at which the luminosity changes sign is the trapping radius
$r_{\rm trap}$.  The lower panel shows the variation of $r_{\rm trap}$
with $\dot{m}$ for a series of models. Except for small values of
$\dot{m}$, where the inner absorbing boundary has an effect, we see
that $r_{\rm trap}$ is quite accurately proportional to $\dot{m}$
\citep{begelman79}. This indicates that HEROIC captures radiation
trapping quite well. In the case of large values of $\dot{m}$, when
the trapping radius is quite far out, the Doppler shifts that describe
radiation advection are fairly small, but the code has no difficulty.

\begin{figure}
\begin{center}
\includegraphics[width=0.5\textwidth]{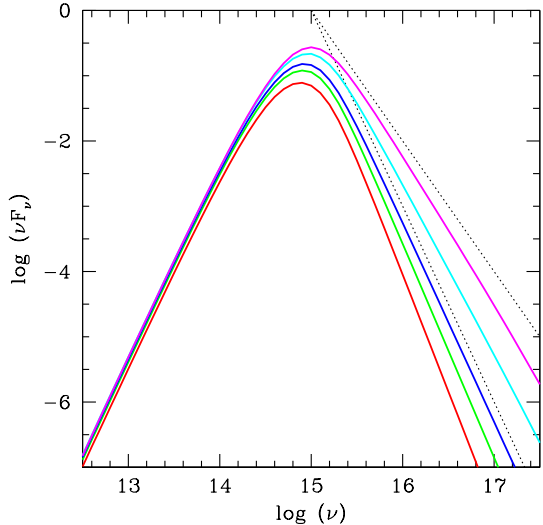}\vspace{-1 in}
\caption{Spectra of escaping radiation for the Bondi accretion problem
  (Fig.~\ref{fig:rtrap}) for $\dot m = 1$ (red), 2 (green), 3 (blue),
  6 (cyan), 10 (magenta). Note the power-law tails at large
  frequencies (in contrast to the spectra in the bottom panel of
  Fig.~\ref{fig:gravitationalredshift}). These power-laws are produced
  by bulk Comptonization. The two dotted lines correspond to photon
  indices of 4 (upper) and 5 (lower).}
\label{fig:bulk_compt}
\end{center}
\end{figure}

Radiation trapping is a $v/c$ effect. A more difficult effect to
capture is bulk Comptonization, which occurs when radiation propagates
through a converging flow (like the Bondi flow).  Figure
\ref{fig:bulk_compt} shows spectra of escaping radiation as calculated
with HEROIC using the same setup as described above. For these
calculations it is assumed that the accreting gas has a temperature of
$10^4$\,K so that there is virtually no thermal Comptonization. In the
absence of bulk Comptonization the spectrum of the escaping radius
should be a perfect blackbody, as in Figure
\ref{fig:gravitationalredshift}. What we see instead are spectra that
are blackbody-like at frequencies below the peak, but are distinctly
power-law in shape at higher frequencies. The high energy power-law is
the result of bulk Comptonization. However, the slopes appear to be
too steep; the photon indices $\Gamma$ we obtain with HEROIC are
larger than those found by \cite{turolla02} by $\Delta\Gamma\sim1$

One reason for the discrepancy could be that HEROIC assumes isotropic
scattering, whereas a correct treatment of the problem should include
the anisotropic nature of Thomson scattering\footnote{Isotropy is
  built into our formula for the source function, which is written in
  terms of only the zeroth angular moment $J_\nu$ of the radiation. It
  is relatively straightforward to include higher-order moments like
  $H_\nu$ and $K_\nu$, since these moments are available during the
  iterations. This will be considered in future upgrades of the
  code.}. A second likely reason is that bulk Comptonization, being a
$(v/c)^2$ effect, is fairly sensitive to how the anisotropic velocity
field in the vicinity of a fluid element is treated. The current
version of HEROIC uses simple linear interpolation between neighboring
cells.  We suspect that interpolation will need to be done more
carefully before we can reliably model bulk Comptonization.

\section{Applications to Accretion Discs}\label{sec:disc}

Much of our information on astrophysical black holes comes from
observations of their accretion discs.  Through modelling the X-ray
continuum of accreting stellar-mass black holes in X-ray binaries, it
is possible to deduce the structure of the accretion disc.
%(e.g., \citealt{shakura_sunyaev73,novikov_thorne73}). 
This can then serve as an indirect probe of the physical properties of
the black hole \citep{mcclintock06,mcclintock14}. In the case of
supermassive black holes, the vast majority of our information comes
from observations of their accretion disc spectra, combined with
efforts to model the observations \citep{krolik99,koratkar99}.

Spectral modeling of black hole accretion discs is complicated by the
fact that the accreting gas is often hot and scattering-dominated.  As
a result, the emerging radiation from the surface of the disc tends to
be significantly ``diluted'' relative to a blackbody spectrum with the
same color temperature. This effect is usually expressed in terms of a
color correction factor $f$ defined by
\begin{equation}
T_{\rm col} = f\, T_{\rm eff}, \label{eq:colorfactor}
\end{equation}
where $T_{\rm col}$ is the color temperature of the emitted radiation
and $T_{\rm eff}$ is the effective temperature corresponding to the
local disc flux.  A key issue then is the estimation of $f$, since it
determines the shape of the resultant disc spectrum.  The earliest
models typically assumed a constant $f\sim1.5$
\citep{mitsuda84,zhang97}.  More recently, much effort has gone
towards pinning down $f$ in the case of X-ray binaries
\citep{shimura95a,shimura95b,merloni00,davishubeny06} via
sophisticated radiative transfer calculations.  In fact, some of the
more recent models go beyond a single number $f$ and estimate in
detail the complete spectrum of the radiation emerging at each radius
of the disc \citep{davishubeny06,davis05,davis06,davis07}.

However, all this prior work suffers from a major limitation: it is
based on plane parallel atmosphere models. Disc coronae are almost
certainly affected strongly by multi-dimensional radiation transfer,
and even disc photospheres are expected to be somewhat affected
(except perhaps in the geometrically thinnest discs). Monte Carlo
methods are very effective for studying multi-dimensional
Comptonization in optically thin regions such as the corona (e.g.,
\citealt{davis09,kawashima12,schnittman13,schnittman13a}), but they
are less useful below the photosphere, where the bulk of the optically
thick radiation is generated. This is where we expect HEROIC to be
useful, since the code seamlessly straddles the optically thick/thin
divide and is inherently multi-dimensional.

Here we present first results from HEROIC. The radiation solutions
described below are multi-dimensional, account self-consistently for
Comptonization, and include all relativistic effects. The intent here
is merely to demonstrate that the code can handle real data taken from
GRMHD simulations of discs.  More detailed discussion is left to
future work.

\subsection{Thin Accretion Disc}

As a model of a thin accretion disc we use one of the GRMHD thin disc
simulations described in \citet{penna10}, which was run using the code
HARM \citep{gammie03,mckinney09}. The simulation did not include
radiation but used of an artificial cooling prescription
\citep{shafee08b,penna10} to keep the disc geometrically thin. From
the HARM simulation we obtain the gas density, velocity and viscous
heating rate in the accretion disc.  We average simulation quantities
over time and azimuth and dimensionalize them for a $10 M_\odot$ black
hole accreting at $\sim60$\% Eddington.  The latter value is chosen to
be consistent with the vertical thickness $h/r\sim0.1$ of the simulated
model. We also extract an integrated luminosity profile from the GRMHD
simulation using the same technique as described in \citet{zhu12}.

While feeding the simulation data into HEROIC, we rebin the data on a
new grid with $n_r = 44$ logarithmically-spaced points in radius, going
from $r=2.17M$ to $r=50.3M$, and $n_\theta = 64$ points in $\theta$,
spaced non-uniformly in such a manner as to preserve the same $\theta$
structure as in the original GRMHD grid (i.e., many more points near
the equator, where most of the gas is located, than near the
poles). The GRMHD data give the vertically integrated energy
dissipation rate per unit disc area. To convert this to heating rate per
unit volume, we arbitrarily assume that heating is proportional to
density,
\begin{equation}\label{eq:qplus}
Q^+ \propto \rho,
\end{equation} 
and present results corresponding to this ansatz.\footnote{We also
  tried models with $Q^+\propto \rho^2, ~\rho^3$. As expected, the
  latter prescriptions put more heating inside the optically thick
  disc and less in the optically thin corona.}  For the gas
temperature, we initially set $T=3\times10^6$\,K in all cells and let
HEROIC solve for the temperature.

While running HEROIC, we set both the inner and outer radial grid
boundaries to have pure outflow conditions for radiation (i.e., no
incident radiation), and the poles to have reflecting boundary
conditions (to account for axisymmetry).  We use $n_A=80$ rays in
angle and set the frequency resolution to be 10 points per decade over
the range $\nu = 10^{16}-10^{20.5}$\,Hz.  For the opacities, we
assume free-free absorption and Thomson scattering,
\begin{equation}\label{eq:alpha}
\alpha_\nu = 1.34 \times
10^{56}\, T^{-1/2} \rho^2\nu^{-3}
(1-e^{-h\nu/kT}) \ {\rm cm^{-1}},
\end{equation}
\begin{equation}\label{eq:sigma}
\sigma_\nu  = 0.4\, \rho \ {\rm cm^{-1}},
\end{equation}
where $T$ is in K, $\rho$ is in ${\rm g\,cm^{-3}}$, and $\nu$ is in
Hz.  The models considered here have typically $T\sim10^7$\,K and
$\rho\sim10^{-4} {\rm g\,cm^{-3}}$ in the disc interior, so scattering
dominates over absorption by a factor of several tens.  Hence we
expect the escaping radiation to exhibit spectral hardening ($f_{\rm
  col} > 1$).  Also, the vertical optical depth through the disc is
quite large, and consequently so is the optical depth across a single
cell in the disc interior. Therefore, it is crucial to include ALI
(\S\ref{sec:ALI}) while converging to the solution. Typically, we need
about 1000 iterations for convergence, though we obtain a fairly good
solution already after a few hundred iterations.

\begin{figure*}
\includegraphics[width=1.05\columnwidth]{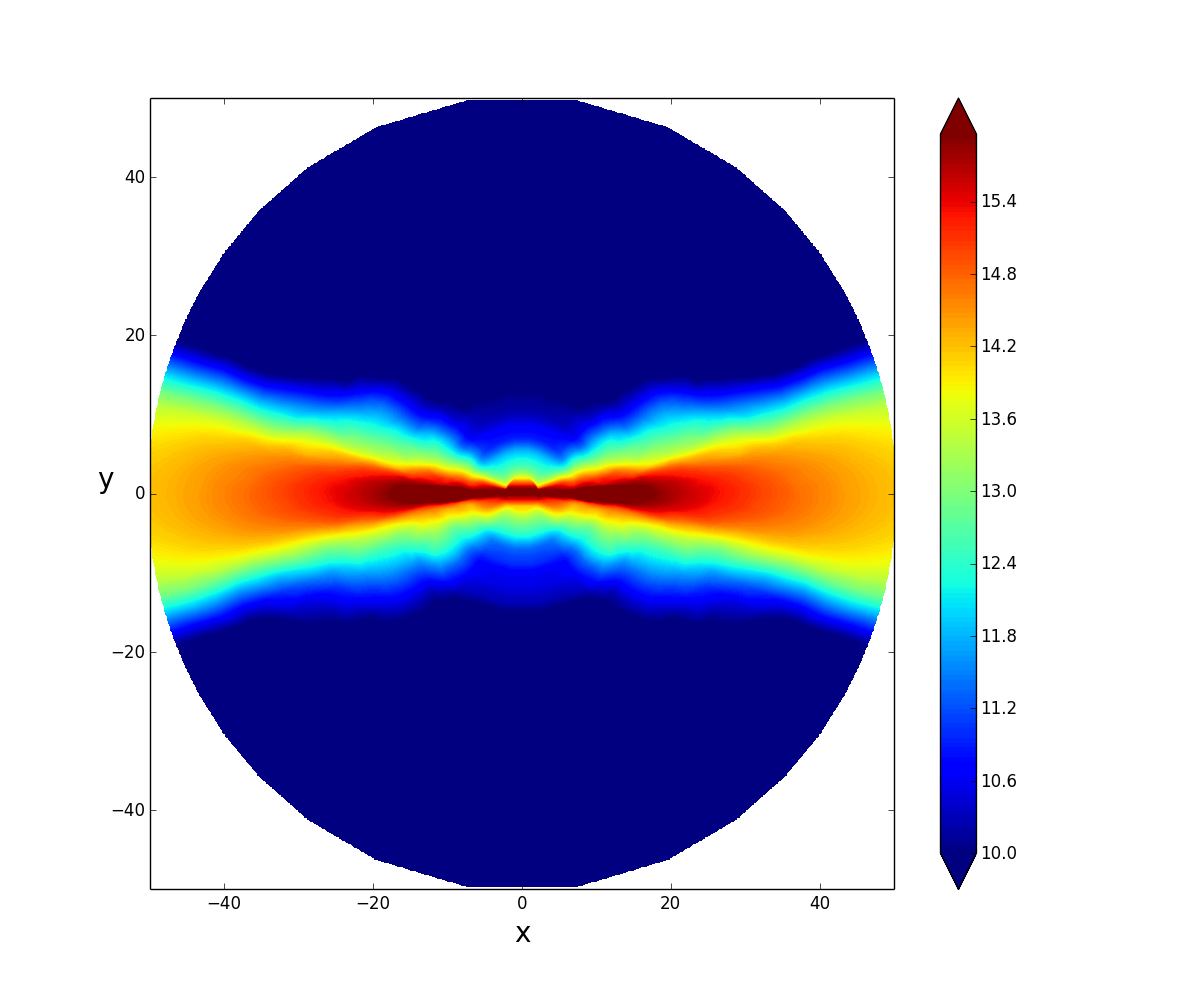}\hspace{-.5cm}
\includegraphics[width=1.05\columnwidth]{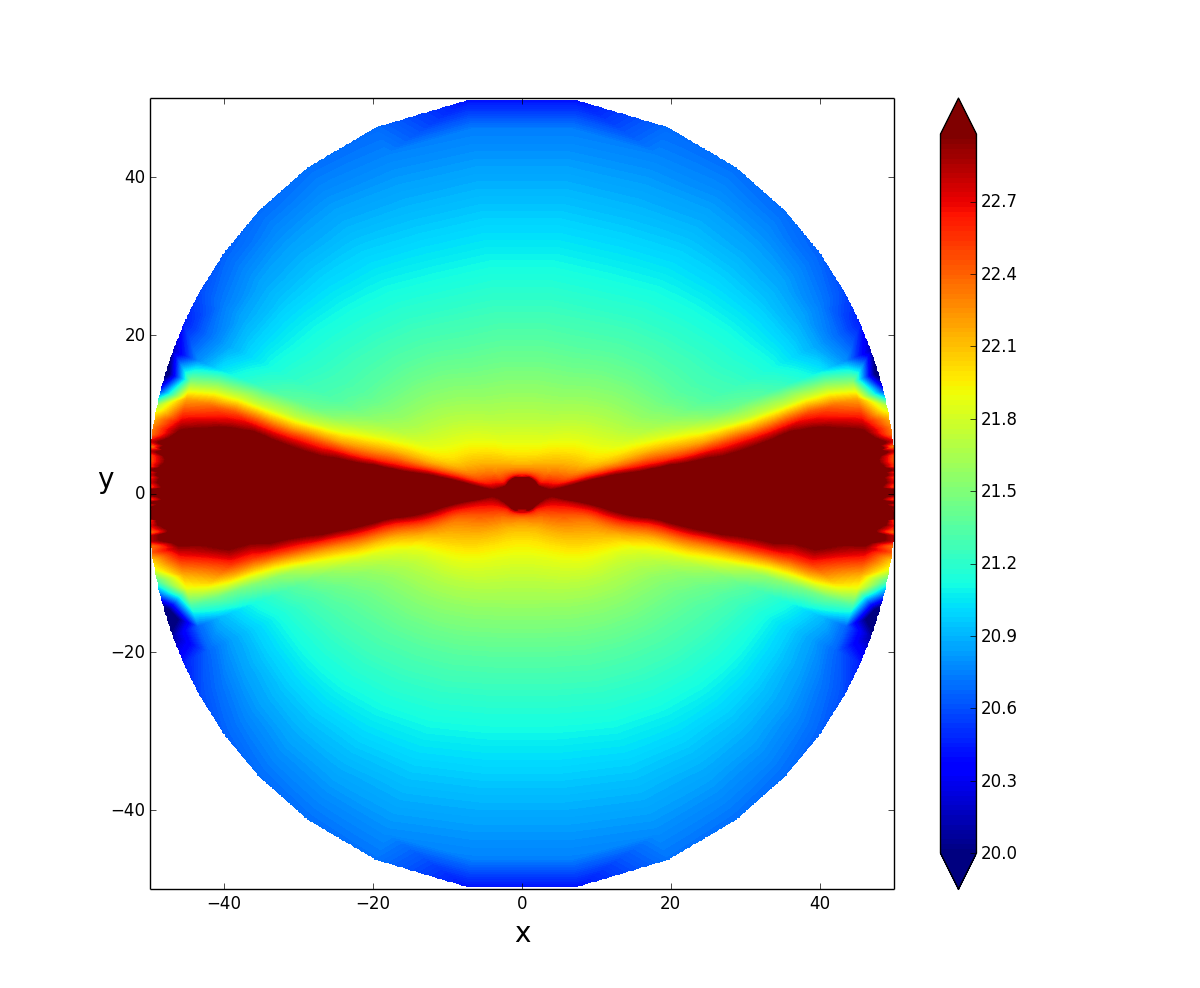} \\
\vspace{-0.25cm}
\includegraphics[width=1.05\columnwidth]{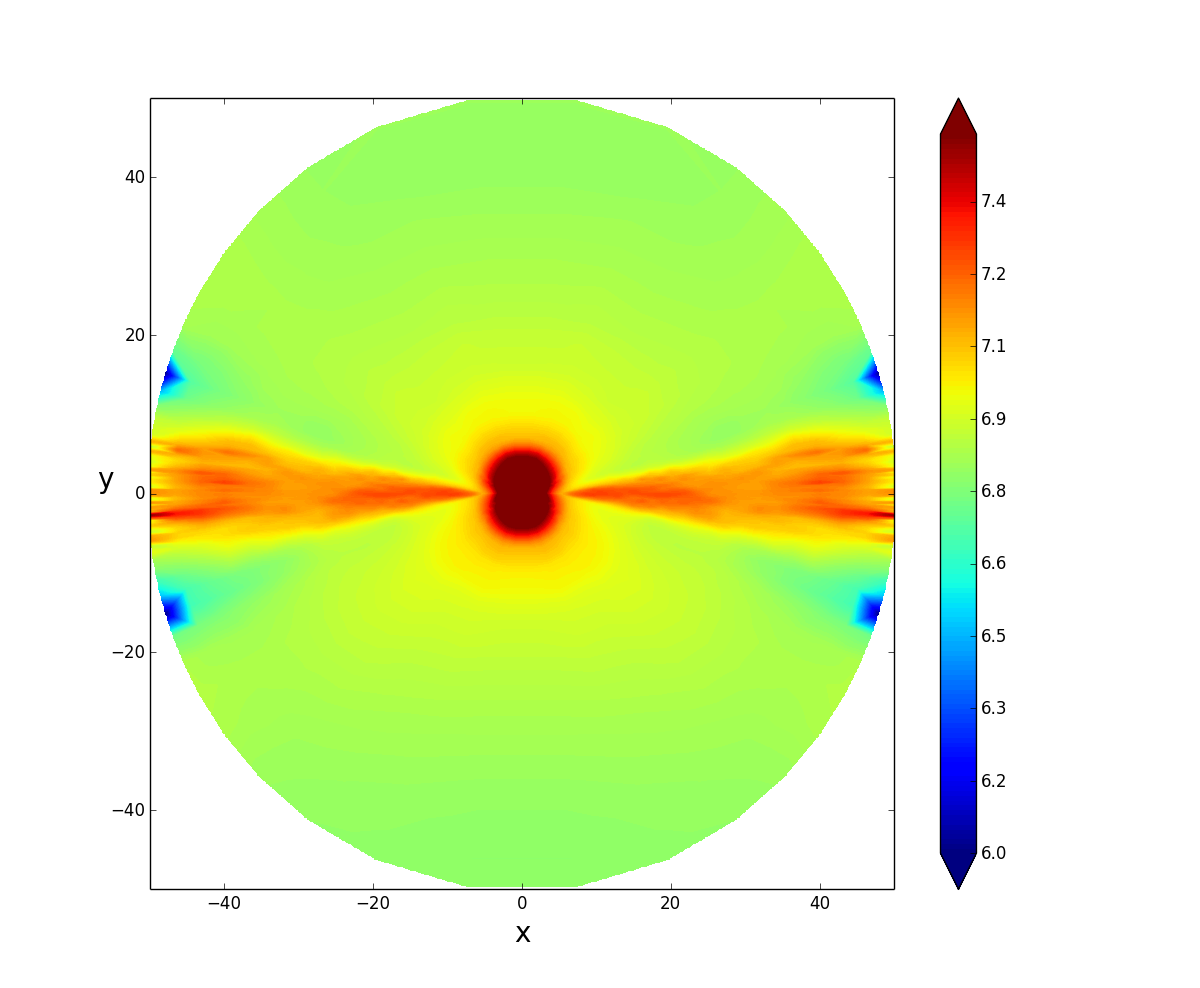}\hspace{-.5cm}
\includegraphics[width=1.05\columnwidth]{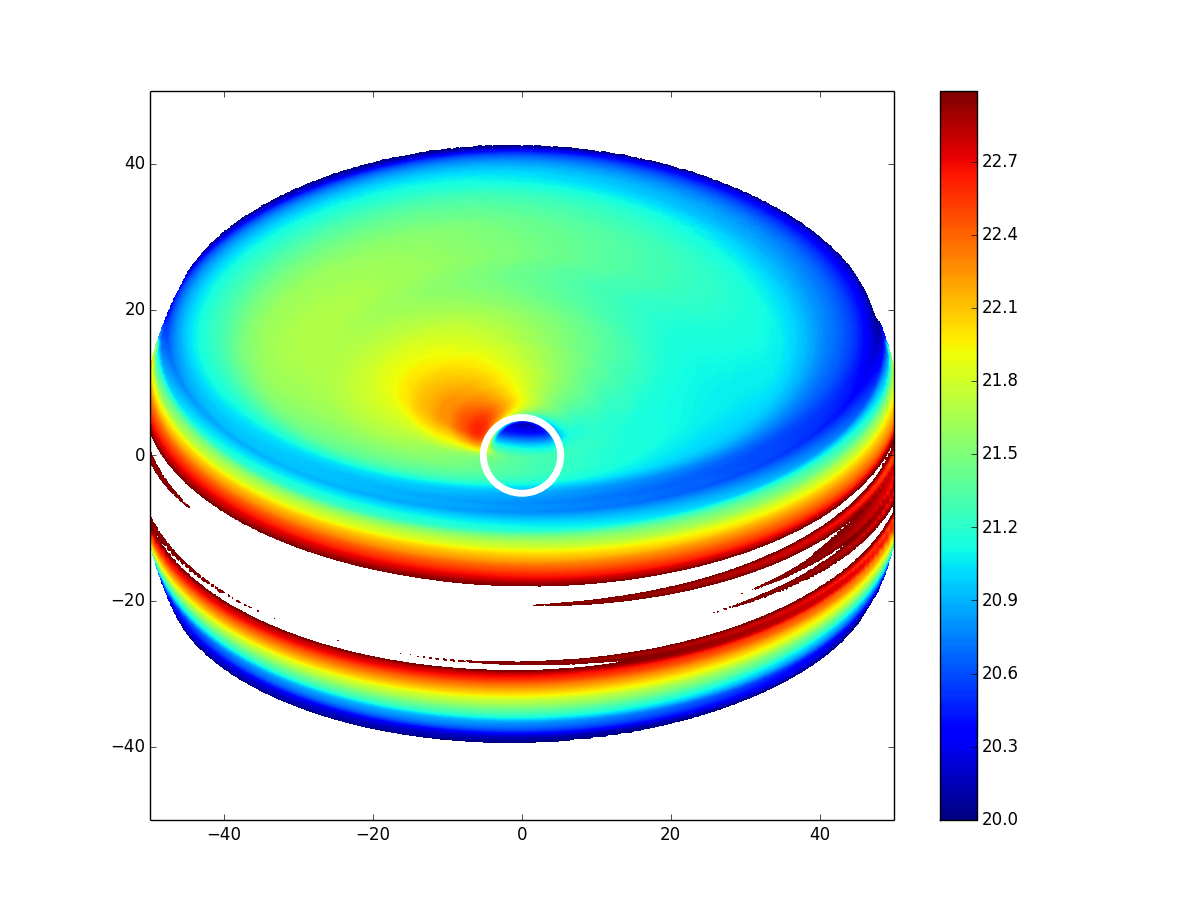} \\
\vspace{-0.25cm}
\includegraphics[width=1.05\columnwidth]{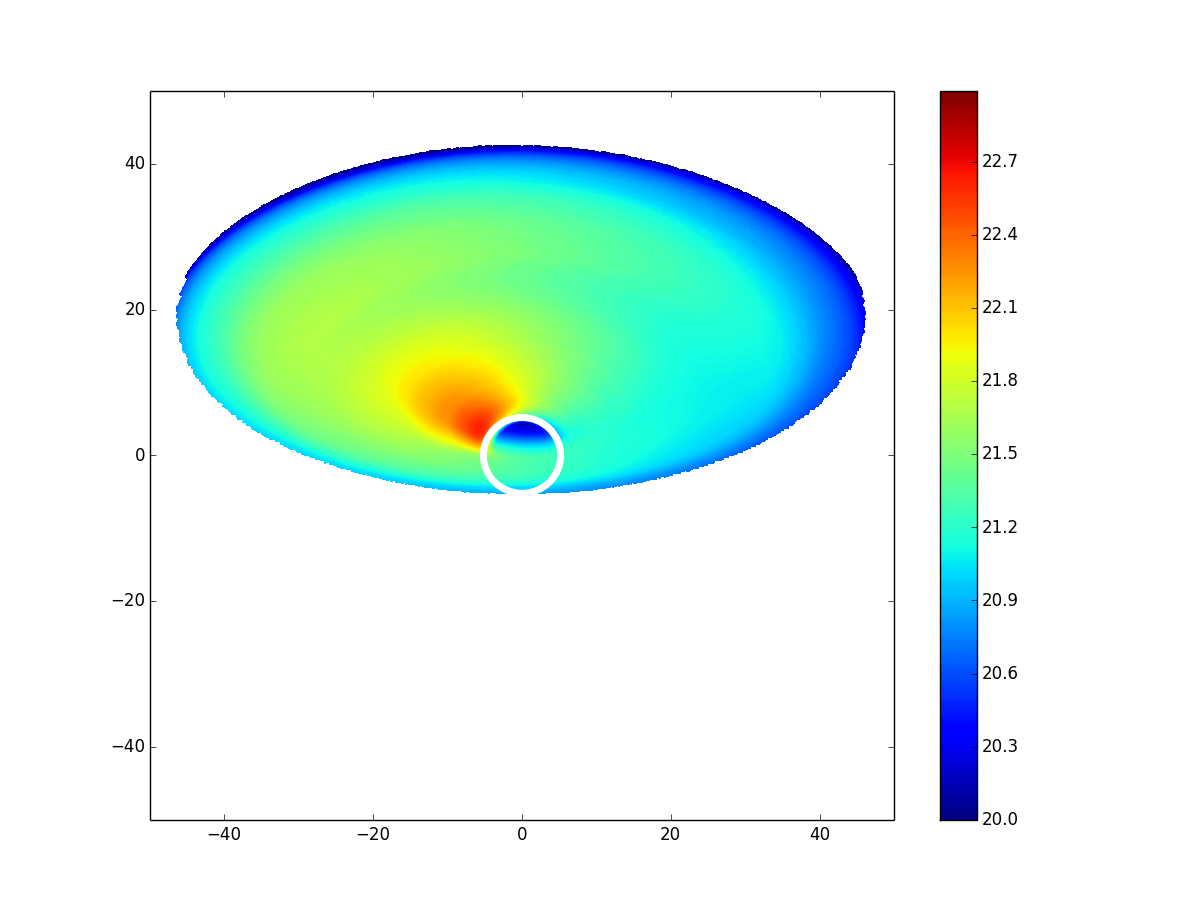}\hspace{-.5cm}
\includegraphics[width=1.05\columnwidth]{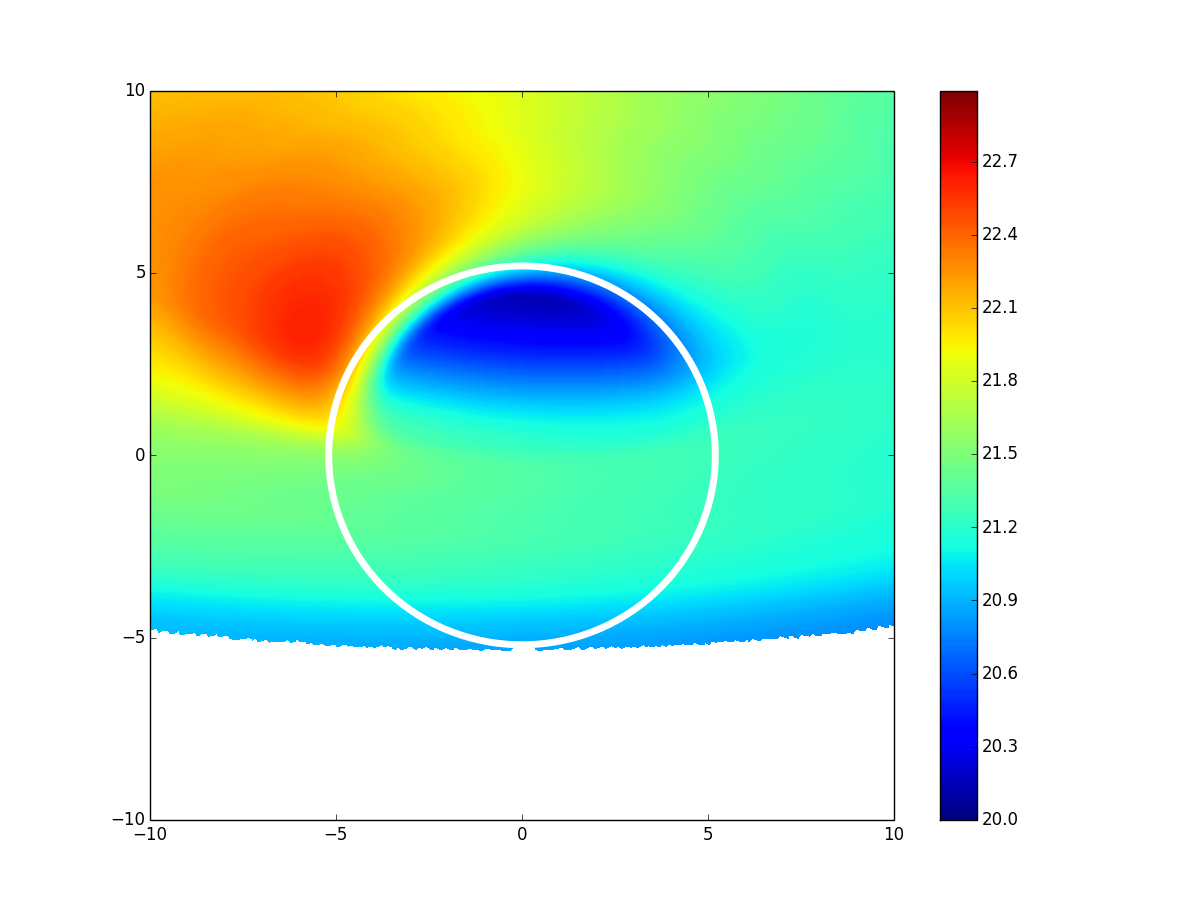} \\
\caption{Application of HEROIC to a GRMHD simulation run with HARM of
  a thin accretion disc around a non-spinning BH. Top Left: Viscous
  heating rate as a function of position in the poloidal plane, as
  estimated from the original simulation (see Eq.~\ref{eq:qplus}). The
  BH is located at $x=y=0$ and the disc mid-plane is oriented
  horizontally. Top right: Frequency-integrated mean radiation
  intensity in the converged HEROIC solution. Center left: Gas
  temperature distribution in the HEROIC solution. Center right:
  Frequency-integrated ray-traced image of the HEROIC solution for an
  observer located at an inclination angle of 60 degrees. The white
  circle corresponds to the apparent size of the photon orbit. Bottom
  left: Same as the previous panel, but restricted to regions of the
  solution whose scattering photospheres lie inside $0.9\, r_{\rm
    max}$. Bottom right: Close-up of the previous panel.}
\label{fig:thindiskimages}
\end{figure*}

Figure \ref{fig:thindiskimages} shows results obtained with HEROIC for
a GRMHD simulation \citep{penna10,zhu12} of a thin accretion disc
around a non-spinning black hole ($a_* = 0$).  The top left panel
shows the heating rate per unit volume versus $(r,\theta)$, estimated
from the simulation data using equation (\ref{eq:qplus}).  While
running HEROIC, we ignore advection (since this is a thin disc, and
also because the simulation data do not provide sufficient information
to estimate the level of advection), so we set $Q^- = Q^+$.

The top right panel in Figure \ref{fig:thindiskimages} shows the
distribution of the mean frequency-integrated radiation intensity,
\begin{equation}\label{eq:J}
J = \int J_\nu d\nu,
\end{equation}
in the converged solution from HEROIC.  The most intense radiation is
in the disc interior and has the typical equatorially flattened shape
we expect, though the geometrical thickness is not as small as one
imagines for a thin disc.  Outside the disc is the coronal region
where the radiation field is more spherical in shape.

The middle left panel in Figure \ref{fig:thindiskimages} shows the
temperature solution obtained by HEROIC. The disc interior is hotter
than the disc surface, as needed to transport out the energy generated
in the interior. The spatial variations (streaks) one sees in the
interior temperature appear to be driven by non-uniform heating caused
by density fluctuations (see eq.~\ref{eq:qplus}). The region inside
the ISCO ($r < R_{\rm ISCO} = 6M$) contains the hottest gas,
distributed almost spherically rather than in a disc. Here the density
is low and the gas needs to be hot in order to radiate whatever
heating is present. 
%\DP{(Can this be where our assumption of no
%  advection breaks down? Even though the gas is very tenuous and
%  cannot cool down, it doesn't stay there long enough to heat
%  up. Should we ``approximate'' this by multiplying the heating rate
%  by a factor proportional to the ratio of the radial inflow time to
%  the heating time?)}  \RN{(Dimitrios, in a previous paper Yucong
%  showed that advection might be important here. I plan to deal with
%  this in a future paper.)}  
Moreover, advection is not necessarily
small \citep{zhu12}, though we have chosen to ignore it for the
present application.  Note that, although the simulation certainly has
heating present inside the ISCO (top left panel), in contrast to the
predictions of the standard thin disc model \citep{novikov_thorne73},
the amount of heating is not large (see
\citealt{gammie99,krolik99a,paczynski00,afshordi03,shafee08a,
  noble09,noble11,penna10,kulkarni11,zhu12} for conflicting
discussions on this issue).  Most of the cooling here is by Compton
scattering.

Considering next the coronal region above the disc, we see that it is
almost isothermal, with $T$ slightly less than $10^7$\,K. The
temperature here is set essentially by the Compton temperature of the
escaping radiation, the latter being determined by the emission from
the disc combined with hotter radiation coming from the plunging
region. There is a small layer of slightly cooler gas just above the
disc photosphere. This region is shielded from direct radiation from
the plunging region, but does receive scattered radiation from the
corona. The cool zone is not as cool as in 1D disc atmosphere models,
as we discuss below.

The remaining three panels in Figure \ref{fig:thindiskimages} show
ray-traced images of the converged disc solution as seen by an
observer at inclination angle 60 degrees. The middle right panel
corresponds to all the radiation emerging from the outer edge of the
computational box at $r_{\rm max} = 50M$. The white equatorial band in
this panel is off-scale and corresponds to radiation coming out of the
disc interior. This should be disregarded since the material here is
visible only because we cut the disc at $r=r_{\rm max}$. The lower
left panel shows a more realistic view of the disc. Here we have
included only those parts of the disc whose the photospheres are
located at radii less than $0.9\,r_{\rm max}$. This effectively
eliminates the extraneous disc interior regions in the previous
panel. The image shows the variation of intensity as a function of
radius, with the hottest regions being closest to the center, and there is
clear evidence for Doppler boosting of the gas rotating towards the
observer (to the left of the black hole) compared to the gas moving
away (on the right). The image also shows distortions on the scale of
the apparent photon orbit ($r = \sqrt{27}M$), indicated by the white
circle.  The lower right panel is a closeup which shows the central
regions of the image and the ``shadow'' of the black hole
\citep{bardeen73,luminet79,falcke00}.

\begin{figure}
\begin{center}  
\includegraphics[width=0.5\textwidth]{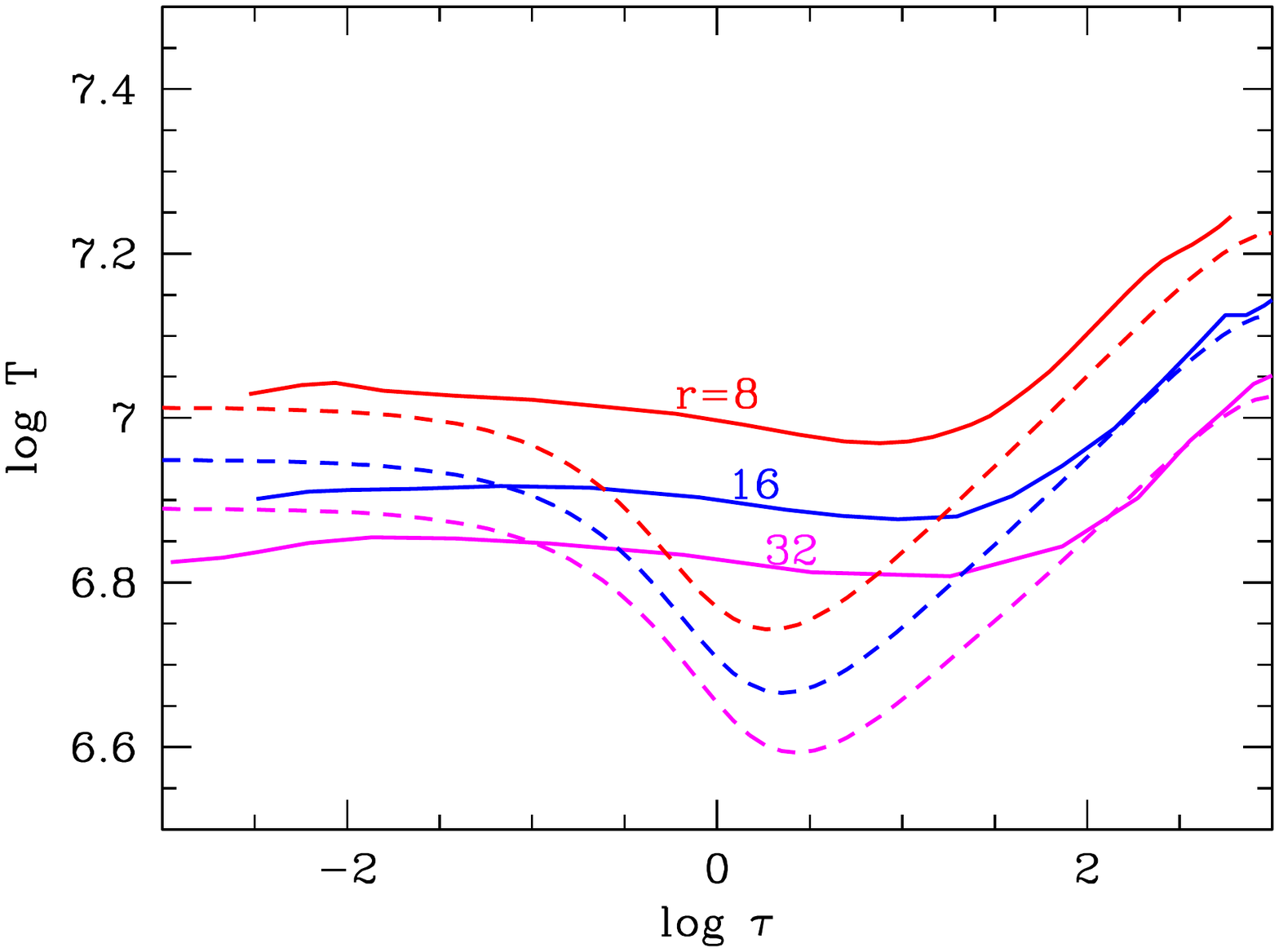}
\vspace{-1.5in}
\caption{Comparison of HEROIC (solid lines) and TLUSTY (dashed lines)
  solutions for the vertical temperature profile at $r = 8$ (red), 16
  (blue), 32 (magenta), for the thin disc model considered in
  Figure~\ref{fig:thindiskimages}.
\label{fig:vertTemp}}
\end{center}  
\end{figure}

In Figure \ref{fig:vertTemp}, we compare the disc temperature profile
as computed by HEROIC with solutions obtained with the 1D radiative
transfer code TLUSTY.\footnote{The TLUSTY solution assumes that disc
  heating is proportional to the density, which is the same as in the
  HEROIC solution presented here.}  In both models, the temperature
profile inside the optically thick disc interior takes on a
characteristic $T \propto \tau^{1/4}$ form, where $\tau$ is the
optical depth from the disc surface.  The primary difference between
TLUSTY and HEROIC occurs above the disc's effective photosphere. The
temperature profile in HEROIC tends towards isothermality whereas the
TLUSTY solution has a pronounced temperature dip at the surface.  This
difference is partly due to the 3D propagation of radiation,
specifically, the effect of gravitationally lensed returning radiation
from the disc and scattered radiation from the corona.  The incoming
flux can penetrate downwards through the cool photospheric surface and
drive the gas towards isothermality via Compton heating.  Some
differences between the 3D GRMHD result and the 1D TLUSTY calculation
are also due to differences in the mass distribution within the disc.
TLUSTY computes the vertical structure of the disc via the condition
of hydrostatic equilibrium, but accounting only for gas and radiation
pressure.  HEROIC does not solve for the vertical density structure,
but takes it from the GRMHD simulation. In the latter, there is
substantial magnetic pressure support and this tends to puff up the
disc, producing structures with more mass at larger height.

The very large difference between the HEROIC and TLUSTY temperature
profiles near the disc photosphere is likely to have observational
consequences. For instance, models of AGN spectra based on plane
parallel atmospheres have a difficult time explaining the absence of
Lyman and other edges in observed spectra \citep{koratkar99}. The flat
temperature profile with depth predicted by HEROIC might provide an
explanation (Shane Davis, private communication).

\begin{figure}
\includegraphics[width=1.0\columnwidth]{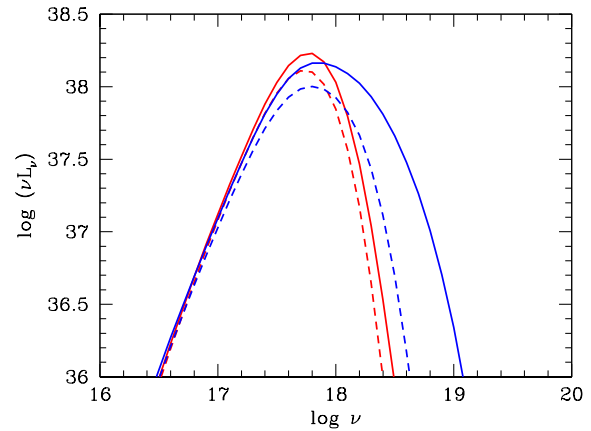}\vspace{-1.5 in}
\caption{Solid red line: Spectrum as seen by an observer at
  inclination angle 60 degrees for a GRMHD thin accretion disc around
  a non-spinning BH (the same model considered in Figures
  \ref{fig:thindiskimages} and \ref{fig:vertTemp}). Dashed red line:
  Spectrum when the viscous dissipation rate is assumed to be the
  analytical result from the \citet{novikov_thorne73} model.  Blue
  solid line: Spectrum corresponding to a GRMHD model of a thin disc
  around a spinning BH with $a_*=0.9$. Dashed blue line: Spectrum when
  the heating rate in this model is set equal to the prediction of
  thin disc theory.}
\label{fig:a0_a9_spectrum}
\end{figure}

%\begin{figure}
%\begin{center}  

%\includegraphics[width=0.5\textwidth]{a0-spec-r20-Compare.pdf}
%\includegraphics[width=0.5\textwidth]{a0-spec-r20-i60-Compare.pdf}

%\caption{Comparison of integrated disc spectra for the inner disc ($r<20$) as computed by HEROIC for the three accretion disc models.  The top panel shows spectra for a the face-on observer ($i=0$), whereas the bottom panel shows spectra for an inclined  ($i=60^o$) observer.
%\label{fig:specCompare}}
%\end{center}  
%\end{figure}  

Figure \ref{fig:a0_a9_spectrum} shows spectra as seen by an observer
at an inclination angle of 60 degrees for two models: (i) the GRMHD
simulation with a non-spinning BH discussed so far (red lines), and
(ii) an equivalent simulation for a spinning BH with $a_* = a/M = 0.9$
\citep{zhu12} (blue lines).  These spectra are computed using the
ray-tracing code for an observation plane located at $r=10^5$.
Starting with the non-spinning BH model, the solid red line
corresponds to the model discussed previously, while the dashed red
line corresponds to a model in which $Q^+$ is set equal to the
analytical prediction of the \citet{novikov_thorne73} model for the
same accretion rate. The latter has no viscous heating inside the
ISCO, whereas the former does (the heating rates outside the ISCO are
also modestly different because of non-zero stress at the ISCO). The
GRMHD model thus predicts a larger luminosity than the equivalent
\citet{novikov_thorne73} model (see
\citealt{penna10,kulkarni11,noble11,zhu12}). However, the shape of the
spectrum is very similar, suggesting that efforts to measure BH spin
by fitting the continuum spectrum of the disc are likely to be
reasonably accurate \citep{mcclintock14}. The spectra corresponding to
the spin 0.9 GRMHD simulation (blue lines) are noticeable hotter for a
comparable luminosity, as expected from thin disc theory, and this
again validates efforts to measure BH spin using disc continuum
spectra. However, here the solid and dashed lines differ more
substantially. The extra dissipation at small radii in the GRMHD
simulation causes the hot gas here to radiate a fair bit, leading to
significantly more emission at high frequencies.

\subsection{Supercritical Accretion Disc}\label{sec:supercriticaldisc}

\begin{figure*}
\includegraphics[width=1.05\columnwidth]{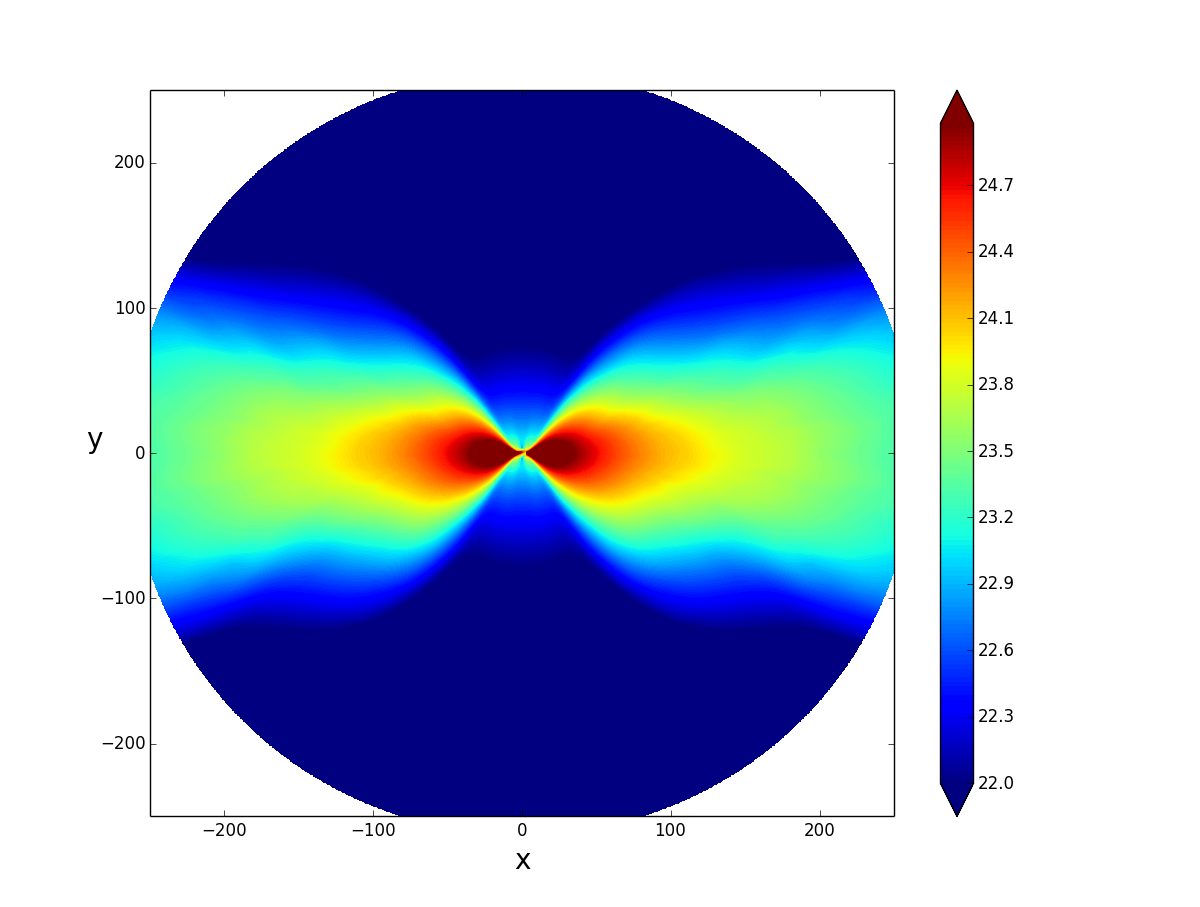}\hspace{-.5cm}
\includegraphics[width=1.05\columnwidth]{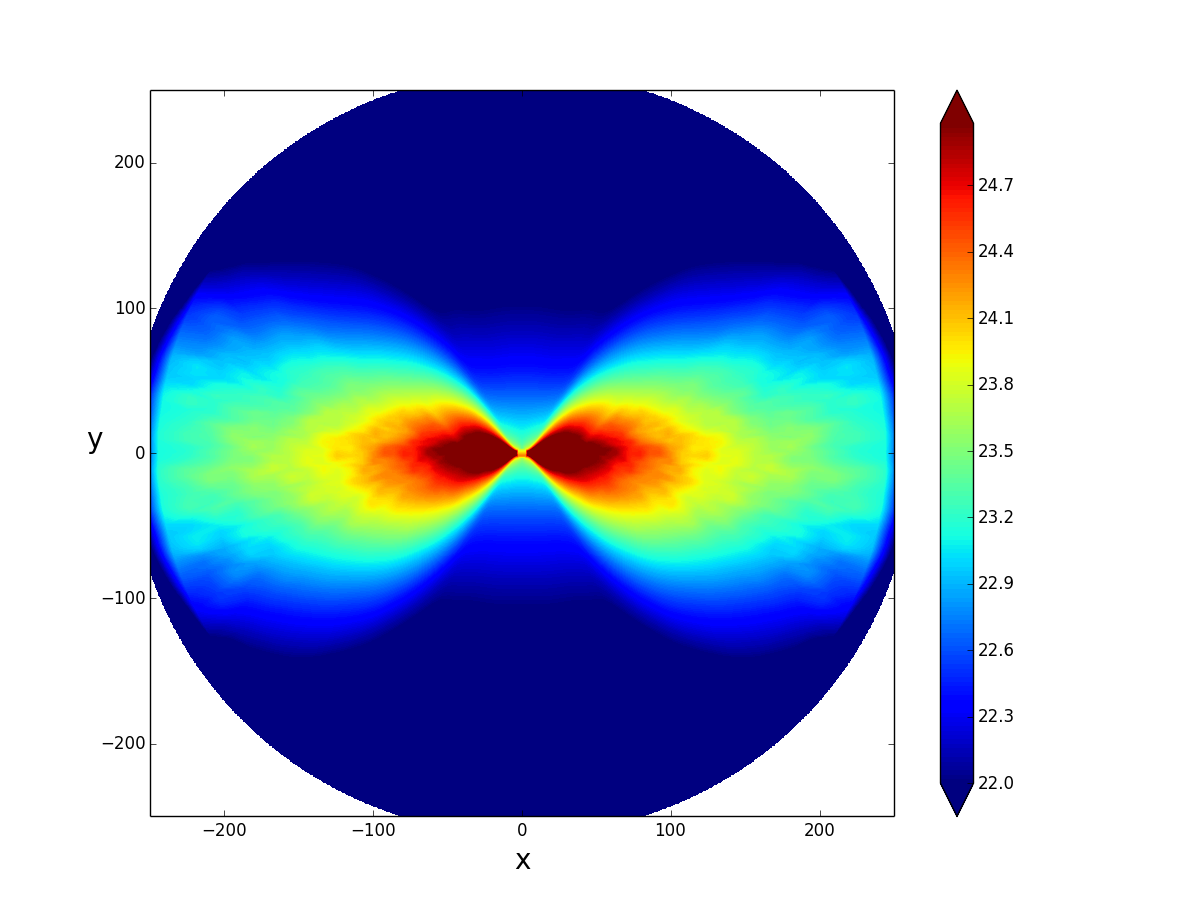} \\
\vspace{-0.25cm}
\includegraphics[width=1.05\columnwidth]{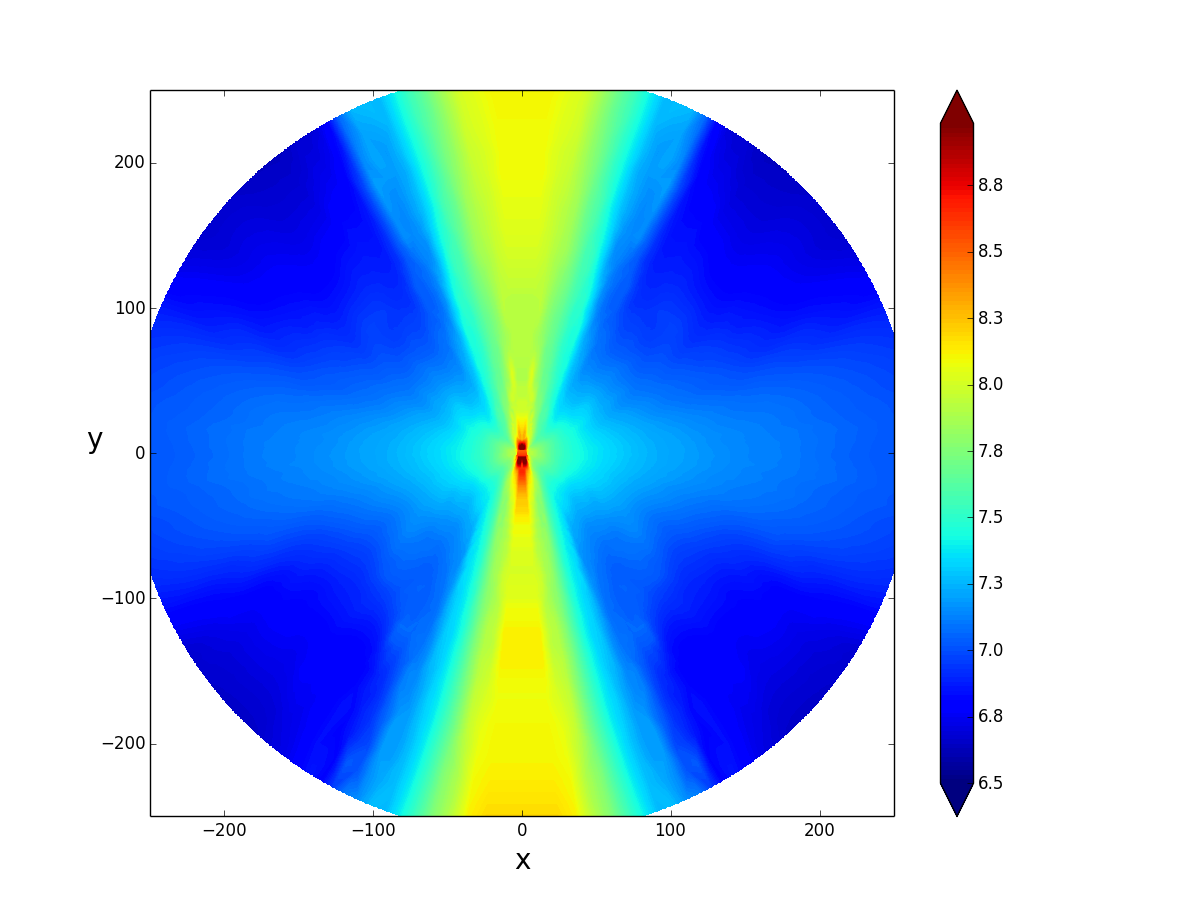}\hspace{-.5cm}
\includegraphics[width=1.05\columnwidth]{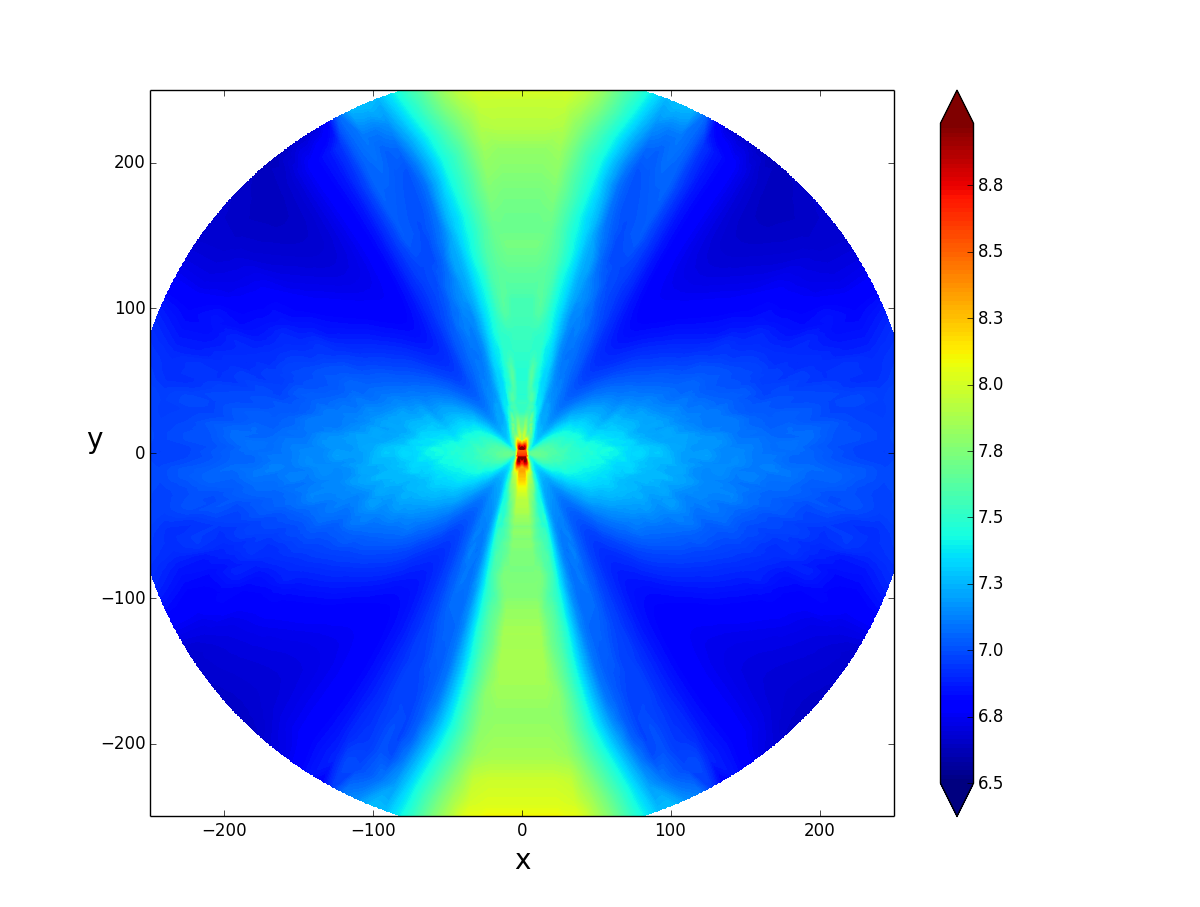} \\
\vspace{-0.25cm}
\includegraphics[width=1.05\columnwidth]{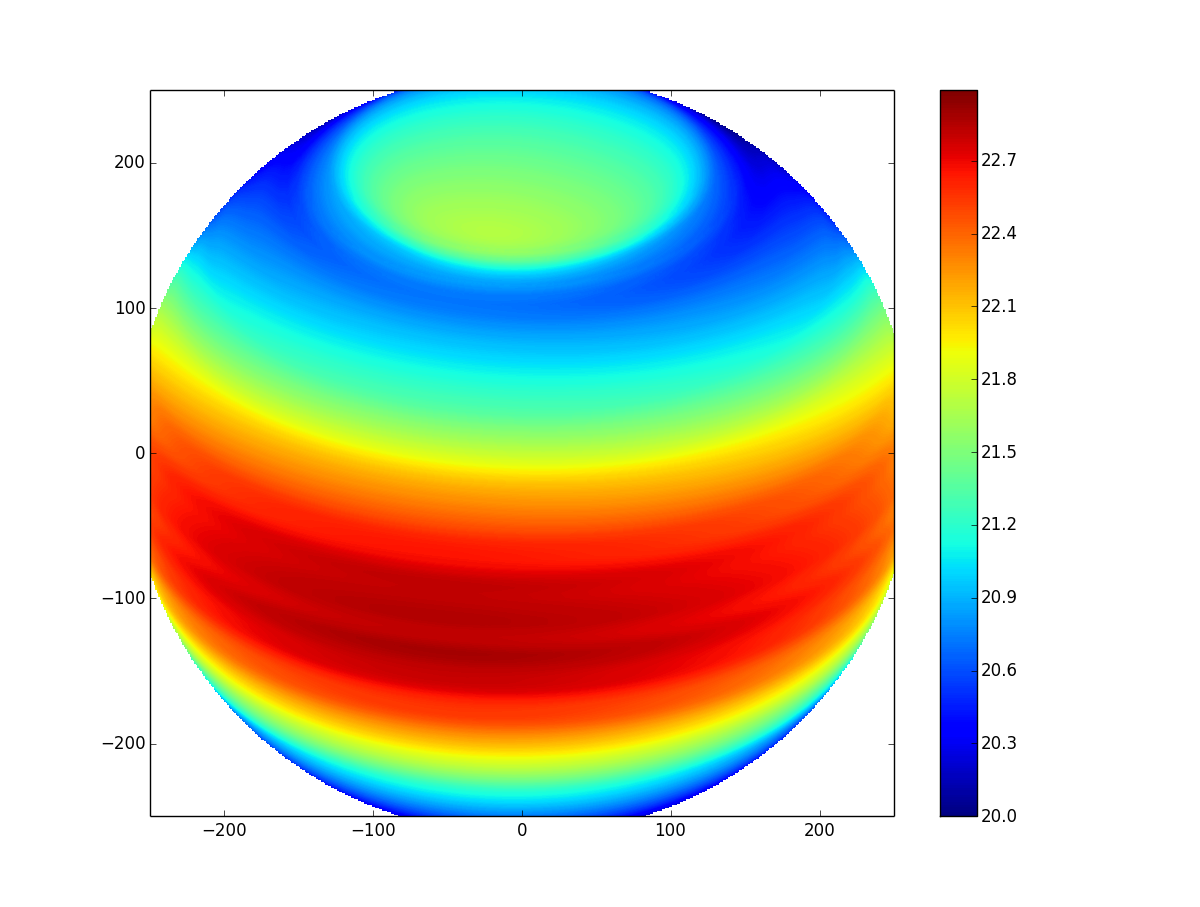}\hspace{-.5cm}
\includegraphics[width=1.05\columnwidth]{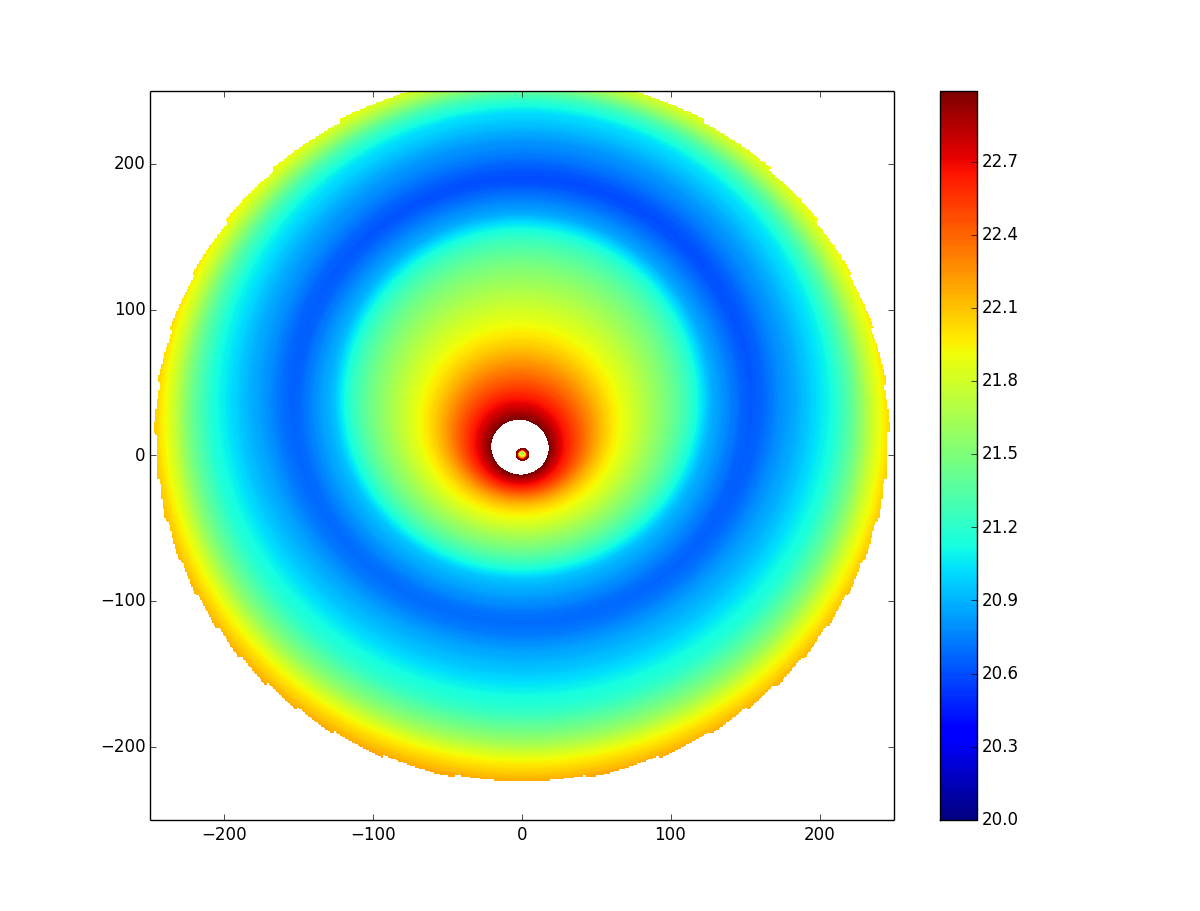} \\
\caption{Application of HEROIC to a GRRMHD simulation run with KORAL
  of a super-Eddington ($\dot{m}=11$) accretion disc around a
  non-spinning BH. The BH is located at $x=y=0$ and the disc mid-plane
  is oriented horizontally. Top left: Frequency-integrated mean
  intensity of the radiation field in the fluid frame as obtained from
  KORAL. Top right: Same quantity as determined from the converged
  HEROIC solution. Center left: Gas temperature distribution from
  KORAL. Center right: Gas temperature distribution in the converged
  HEROIC solution. Bottom left: Frequency-integrated ray-traced image
  of the HEROIC solution for an observer located at an inclination
  angle of 60 degrees. Bottom right: Image for an observer at 10
  degrees, but showing only the regions of the solution at $\theta <
  70$ degrees.}
\label{fig:supercriticaldisk}
\end{figure*}

For our second application we consider a general relativistic
radiation MHD (GRRMHD) simulation of a supercritical ($\dot{m} >
\dot{m}_{\rm Edd}$) accretion flow run with the code KORAL
\citep{sadowski13,sadowski14}. The particular simulation we analyse is
taken from \citet{sadowski15b} and corresponds to a $10M_\odot$ BH
accreting at about 11 times the Eddington rate.  This simulation used
a photon-conserving form of Comptonization. From the time-averaged
simulation data we obtain the density and four-velocity of the fluid.
These quantities are input into HEROIC on a grid going from $r=3.4$ to
$r=250$. Since KORAL evolves the radiation field, it obtains directly
the cooling rate $Q^-$ as a function of position. From this, we
calculate $Q^+$ via equation (\ref{eq:qadv}); in this step, we use the
KORAL temperature for consistency. However, once we input these $Q^+$
values into HEROIC and as the iterations proceed, we use the
HEROIC-derived temperatures to estimate $Q^-$, again through equation
(\ref{eq:qadv}). HEROIC is run with 80 ray angles and 61 frequencies
distributed uniformly in $\log\nu$ between $\nu=10^{16}$\,Hz and
$\nu=10^{22}$\,Hz. For the opacities we use equations (\ref{eq:alpha})
and (\ref{eq:sigma}).

Figure \ref{fig:supercriticaldisk} shows the results. The top left
panel shows the mean intensity $J$ (see eq.~\ref{eq:J}) obtained in
the KORAL simulation. Since this is a super-Eddington flow, the
optically thick regions of the disc are geometrically thick.  Along
the poles are two moderately wide funnels which are optically
thin. Radiation from the main disc flows into the funnel and escapes
in twin beams, while at the same time accelerating polar gas in
relativistic jets.  Note that the very central region of the funnel is
somewhat devoid of radiation (the tiny thin vertical drak blue line at
the axis). This is a well-known weakness of the M1 closure scheme on
which KORAL is based \citep{sadowski14}, although the effect is not as
strong as it could be in this particular simulation since an
artificial radiation viscosity was applied \citep{sadowski15a}. The
top right panel in Figure \ref{fig:supercriticaldisk} shows the mean
radiation intensity obtained by HEROIC for the same system. The
solution is very similar to the KORAL solution in the main disc. The
funnel region, however, is smoother than in the previous panel. In
particular, the radiation deficit near the pole is no longer present.
Also, the HEROIC radiation field is a little stronger compared to
KORAL in the funnel.

The middle two panels in Figure \ref{fig:supercriticaldisk} show the
temperature distributions obtained with KORAL (left) and HEROIC
(right). There are some differences in the disc interior, but these
are not important since they do not affect the radiation that reaches
the observer. For the latter, what matters is the temperature in the
optically thin funnel region and at the photosphere. Here we see
fairly good agreement, though HEROIC gives slightly lower temperatures
than KORAL. A likely reason for this is the fact that the radiation
density is different in the two models, as already discussed. Since
the gas temperature in the funnel is largely determined by Compton
scattering, it is natural for the HEROIC model with its larger
radiation energy density to be cooler. It should also be kept in mind
that KORAL uses a crude representation of the radiation, which
consists of just the bolometric energy density and bolometric flux (4
numbers at each location), and computes the radiative transport with
frequency-integrated effective opacities.  The HEROIC model shown
here, by contrast, solves for the intensities on 80 ray angles, each
over 61 frequencies, and uses a frequency-dependent opacity
(eq.~\ref{eq:alpha}). These are large differences, and it is natural
for the results to deviate.

The bottom panels in Figure \ref{fig:supercriticaldisk} show
ray-traced images. The panel on the left shows all the radiation from
the HEROIC solution as would be seen by an observer at an inclination
angle of 60 degrees. This is not particularly meaningful because most
of the gas is optically very thick and even the optically thin region
in the funnel at the top would not be visible if we had not truncated
the model at $r_{\rm max} = 250$. Nevertheless, the similarity to the
bottom left panel in Figure 9 of \citet{ohsuga05} is striking, despite
the many differences in the methods used in the two studies. The right
panel in Figure \ref{fig:supercriticaldisk} shows the view of the
region of the solution at $\theta<70$ degrees for an observer at a low
inclination angle of 10 degrees. This observer, and others up to an
inclination angle of around 25 degrees, will be able to see the walls
of the funnel (the circular central cyan-green-yellow-red region) and
down near the BH horizon (maroon-white). In the image of a real
system, this funnel region would be surrounded by the photosphere of
the rest of the optically thick disc. However, the KORAL simulation
under consideration did not extend over a large enough volume to
reliably model the photosphere. Therefore, the outer half of the image
shown in this panel (starting with dark blue, extending to
cyan-green-yellow) should not be taken too seriously.

\begin{figure}
\includegraphics[width=1.0\columnwidth]{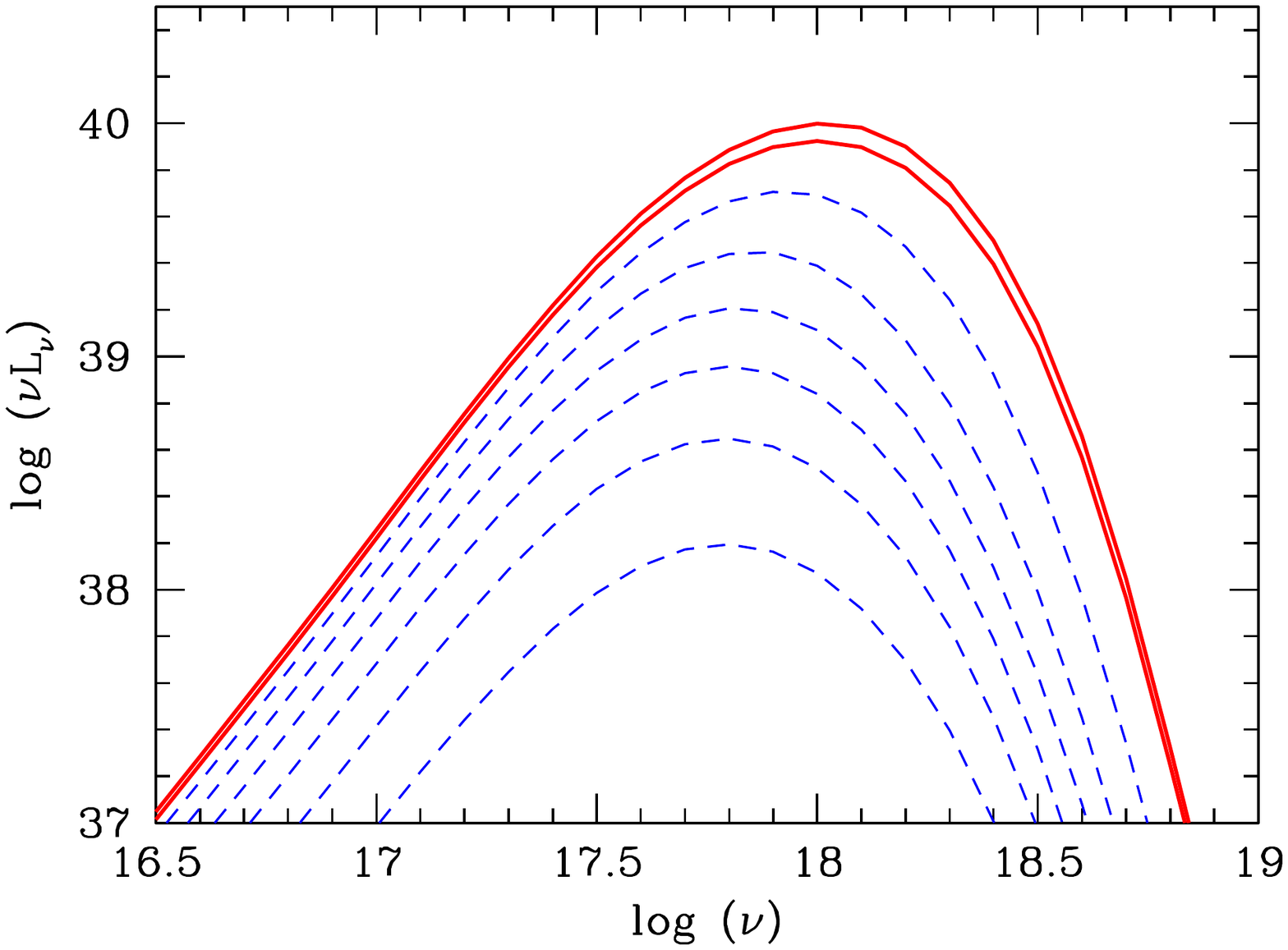}\vspace{-1.8 in}
\includegraphics[width=1.0\columnwidth]{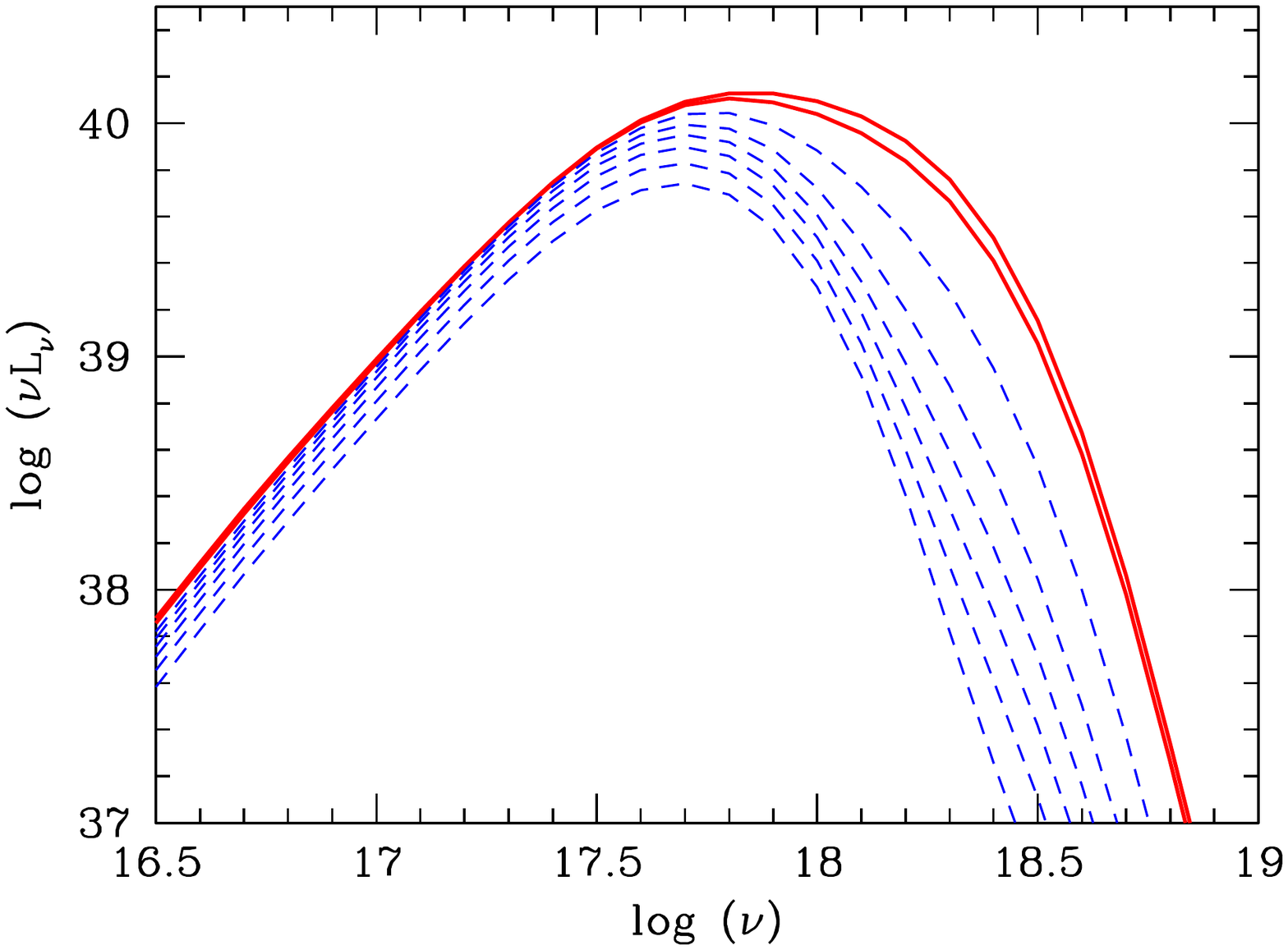}\vspace{-1.5 in}
\caption{Top panel: Spectra corresponding to radiation from the funnel
  region of the super-Eddington BH accretion model considered in
  Fig.~\ref{fig:supercriticaldisk}. Solid red lines: Spectra as seen
  by observers at inclination angle (from above) 10 and 20 degrees,
  respectively. Dashed blue lines: Spectra for observers at (from
  above) 30, 40, 50, 60, 70, 80 degrees, respectively.  Lower panel:
  Spectra when the entire region of the model up to $\theta=70$\,deg
  is considered.}
\label{fig:spectrumsupercritical}
\end{figure}

The upper panel in Figure \ref{fig:spectrumsupercritical} shows
spectra of radiation emerging from the funnel as viewed by observers
at various inclination angles. Here, the funnel is defined as any
region for which the scattering photosphere lies below $0.9\,r_{\rm
  max}$. The three red spectra at the top correspond to pole-on
observers who are able to see all the way down to the BH. For these
observers, the object would appear as a bright super-Eddington X-ray
source with peak emission at around 5~keV.  The remaining blue spectra
are for more inclined observers. However, these spectra are less
meaningful since radiation from the funnel escapes in these directions
only because we trucated the disc at $r_{\rm max}$. In a real system,
the funnel would extend farther out in radius and no radiation from
the hot inner regions would go towards the sides.

The spectra in the upper panel in Figure
\ref{fig:spectrumsupercritical} are too peaked and blackbody-like
compared to the spectra obtained by \citet{kawashima12} from analogous
Newtonian radiation-hydrodynamic simulations of super-Eddington
accretion discs. In part this is because they considered also
radiation from the more optically thick wind outside the funnel.
Therefore, for comparison, we show in the lower panel of Figure
\ref{fig:spectrumsupercritical} spectra as seen by distant observers
but now considering the entire HEROIC solution out to polar angle
$\theta = 70$\,deg. These spectra do have more radiation at softer
photon energies and more closely resemble the results shown in
\citet{kawashima12}. However, the HEROIC spectra are a little too
soft, peaking at about 5\,keV rather than at 10\,keV.

\begin{figure}
\includegraphics[width=1.0\columnwidth]{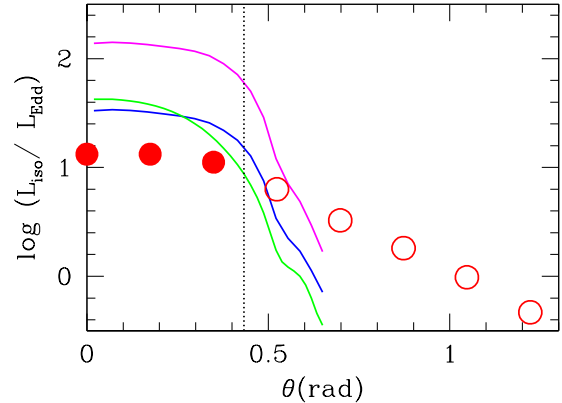}\vspace{-1.5 in}
\caption{Isotropic equivalent luminosity (in Eddington units) of the
  super-Eddington BH accretion model considered in Figures
  \ref{fig:supercriticaldisk} and \ref{fig:spectrumsupercritical}, as
  a function of polar angle $\theta$.  Green line: Luminosity
  estimated from the original KORAL simulation based on the radial
  radiative flux at the outer edge of the box, $r=r_{\rm max}
  =250$. Magenta line: Luminosity estimated at the same radius from
  the HEROIC solution when the gas temperature is held fixed at the
  KORAL values. Blue line: Luminosity from the HEROIC solution when
  the temperature is solved self-consistently within HEROIC. Vertical
  dotted line: Nominal edge of the optically thin funnel region. Red
  symbols: Isotropic equivalent luminosity measured by a distant
  observer as a function of inclination angle. Only radiation emerging
  from inside the funnel region is included. Solid symbols are more
  meaningful since the BH at the base of the funnel is visible to
  these observers.}
\label{fig:luminositiessupercritical}
\end{figure}

Finally, Figure \ref{fig:luminositiessupercritical} shows several
versions of the istropic equivalent luminosity $L_{\rm iso}$ of the
supercritical BH accretion model under consideration as a function of
inclination angle. The various lines are obtained by taking the radial
flux as a function of $\theta$ at the outer edge of the box ($r_{\rm
  max}=250$) and converting it to an effective isotropic equivalent
luminosity. The green line shows the luminosity from the original
KORAL simulation. Here, the funnel region corresponds to $\theta <
0.43$\,rad, indicated by the vertical dotted line. This is the only
part of the radiation that is guaranteed to escape to infinity. The
radiation at larger angles is inside optically thick gas (even at
$r=250$) and it is not clear what fraction of this energy flux will
finally escape as radiation.  

The magenta line in Figure \ref{fig:luminositiessupercritical} shows
the isotropic equivalent luminosity at $r=250$ obtained from the
HEROIC solution when we keep the gas temperature fixed at the KORAL
values and solve self-consistently only for the radiation field. This
curve has qualitatively the correct shape versus $\theta$, but the
luminosity is several times too large. On the other hand, the blue
line shows the result when we solve for both the temperature and the
radiation field with HEROIC (all the previous results in
Figs.~\ref{fig:spectrumsupercritical} and \ref{fig:supercriticaldisk}
correspond to this solution). Now we see much closer agreement with
the KORAL result. The HEROIC profile is slightly more flat-topped, but
the integrated luminosity is close. This comparison highlights the
important point that, when post-processing GRRMHD simulations, it is
necessary to solve self-consistently for the gas temperature, a point
made earlier by \citet{schnittman13a}. It is particularly important
with Comptonization because small changes in the temperature can cause
large changes in the radiation energy density.

The red symbols in Figure \ref{fig:luminositiessupercritical} show the
isotropic equivalent luminosity as measured by distant observers
located at different inclination angles. These are computed from the
HEROIC model corresponding to the blue line, i.e., with both
temperature and radiation field calculated self-consistently.  Only
the results corresponding to small inclination angles ($<25$\,deg,
shown by solid symbols) are meaningful. For these observers, the
apparent luminosity will be highly super-Eddington. In fact, both the
luminosity and the spectrum (Fig.~\ref{fig:spectrumsupercritical}) for
these face-on observers agree qualitatively with observations of
ultra-luminous X-ray sources (e.g., \citealt{miyawaki09,kawashima12}).
Note that, on top of the luminosities computed here, there would be an
additional $\sim1 L_{\rm Edd}$ of radiation from the rest of the
optically thick disc. This is a small correction, but it would have a
softer spectrum and would make noticeable changes to the low-energy
end of the spectra shown in Figure \ref{fig:spectrumsupercritical}.

A notable feature of the red symbols in Figure
\ref{fig:luminositiessupercritical} is that, at small angles, they lie
below the blue line by a factor $>2$. That is, if we estimate the
luminosity at a given $\theta$ in the funnel via the radiative flux at
radius $r_{\rm max}$ (blue line), the result will be an overestimate
compared to the luminosity that an observer at infinity at the same
inclination angle $\theta$ would observe (filled red circle). This is
just a matter of geometry. A good fraction of the radiation at $r_{\rm
  max}$ goes off sideways towards observers at larger inclination
angles, which is why the open circles in Figure
\ref{fig:luminositiessupercritical} lie above the corresponding blue
line. One other relevant comment is that HEROIC assumes a thermal
medium, whereas some of the emission from the jet in the funnel might
be non-thermal. Such a component would appear as a high energy
power-law tail in the spectrum.

\section{Summary and Discussion}\label{sec:discussion}

In this paper we described an extension of our radiative transfer code
HERO (Paper 1) that now enables us to handle Compton scattering.
Given the density, velocity and heating rate of the gas as a function
of position, the new code HEROIC self-consistently solves for both the
radiation field and the gas temperature. The code handles a wide range
of optical depths, from optically very thick to very thin, and
operates in multiple dimensions within a general relativistic
space-time. It is suitable for modeling radiation in high energy
astrophysical objects where thermal Comptonization is important.

We described a number of tests of HEROIC. We showed that the code
reproduces known results for thermal Comptonization, both in one
dimension (plane parallel geometry) and in multiple dimensions
(spherical geometry treated in axisymmetry). It also handles
relativistic effects like Doppler shift, which is important for
modeling radiation trapping and advection, gravitational redshift, and
ray-deflection. In addition, the code produces bulk Comptonization
when there is a converging flow (as in spherical accretion), though
the slope of the power-law tail is too soft.

The inability of HEROIC to model bulk Comptonization accurately is
likely because the code assumes isotropic scattering rather than using
the correct angular distribution. This could be rectified in the
future. Also, only a short characteristics version of HEROIC is
available at the moment; the next step is to develop a long
characteristics version. Neither of these shortcomings is serious, and
both are easily overcome. A more important limitation is that HEROIC
assumes the gas to be thermal and in LTE. Extension to NLTE is
possible, in principle, but will require a major upgrade. Extension to
non-thermal processes would be equally difficult, requiring at the
very least prescriptions for the energy-dependent heating of
non-thermal electrons and for various opacities.

Other limitations are inherent to the very structure of HEROIC and
cannot be overcome.  The code assumes that there is no time dependence
in any quantity, thus it is most appropriate for studying time-averged
properties of objects.  HEROIC could be used to study time-dependent
phenomena, but it would have to be under the ``fast-light''
approximation, where time delays are not taken into account. HEROIC is
a {\it radiation post-processor} which takes density, velocity,
etc. from other codes and obtains a more detailed, and hopefully more
accurate, solution for the radiation field. In the process it also
improves the gas thermodynamics by re-solving for the
temperature. However, there is no dynamics in the code. In principle,
for simple geometries, e.g., plane parallel or spherically symmetric
systems, dynamics could be built into the structure of the code by
including additional conditions (force equation, energy equation), but
it is not clear that this would be an improvement on other simpler
codes.

As examples of how HEROIC might be used, we presented in this paper
two applications. First, we took a GRMHD simulation of a thin
accretion disc around a BH and used HEROIC to solve for the gas
temperature and radiation field. The original GRMHD simulation did not
include radiation but used an artificial cooling
prescription. Therefore, we had to make an educated guess regarding
the distribution of viscous heating in the system.  Given this guess,
HEROIC was successfully able to solve the multi-dimensional radiative
transfer problem. The most striking result was that the temperature
profile determined by HEROIC was noticeably different from that
previously obtained from 1D plane-parallel disc atmosphere models.
This effect could have implications for modeling disc spectra and
deserves to be studied further.

In the second application, we took data from a GRRMHD simulation of a
supercritical accretion disc around a BH (accretion rate of 11
Eddington) and solved for the gas temperature and radiation field
using HEROIC. We confirmed that observers who view such a system from
small inclination angles would see very large apparent luminosities,
up to tens of Eddington. The spectrum could still be quite thermal
(assuming there is no non-thermal heating of electrons) and
blackbody-like, and there might not be any spectral indication for
relativistic beaming. Models like this may explain apparently
super-Eddington objects such as ultra-luminous X-ray sources.

To conclude, we view HEROIC as a tool to bridge the gap between
GRMHD/GRRMHD simulations and observations. The simulations that have
been done to date are highly sophisticated in their treatment of
dynamics, but they are relatively crude in how they handle
radiation. Either they ignore radiation altogether (GRMHD) or, as in
the GRRMHD codes currently available for simulating black hole
accretion \citep{sadowski14, mckinney14, fragile14,
  takahashi_ohsuga15}, they treat radiation effectively as a fluid
described by a few angular moments. More sophisticated techniques have
been applied to treat radiation in Newtonian simulations of discs
\citep{jiang12,jiang14}, but even these methods, while allowing more
angular structure in the radiation field, are generally limited to
frequency-integrated quantities. It is far too expensive to run
multi-dimensional hydrodynamic or MHD simulations and to follow at the
same time at each spatial location many ray directions and many
frequencies. Until computers become much more powerful than they are
today, post-processing of simulation output seems to be the only way.
HEROIC is designed to fill this need.

\section{Acknowledgements}

RN and YZ were supported in part by NSF grant AST1312651. RN also
received partial support under NASA grant TCAN NNX14AB47G.  AS
acknowledges support by NASA through Einstein Postdoctotral Fellowship
number PF4-150126 awarded by the Chandra X-ray Center, which is
operated by the Smithsonian Astrophysical Observatory for NASA under
contract NAS8-03060. AS also thanks the Harvard-Smithsonian Center for
Astrophysics for hospitality.  The authors acknowledge computational
support from NSF via XSEDE resources (grant TG-AST080026N), and from
NASA via the High-End Computing (HEC) Program through the NASA
Advanced Supercomputing (NAS) Division at Ames Research Center.

\appendix

\section{The Fokker-Planck Equation for Comptonization}\label{secA:FP}

In the limit of small changes in the photon energy per scattering,
Comptonization of an isotropic distribution of photons is described by
the Fokker-Planck equation~\citep{barbosa82}
\begin{eqnarray}
\frac{\partial n}{\partial t} &=& \frac{1}{\nu^2} \frac{\partial}{\partial\nu}
\left\{n(1+n) \left[ -\nu^2 \left\langle \frac{\Delta\nu}{\tau} \right\rangle
+\frac{\partial}{\partial\nu}\left( \nu^2 \left\langle \frac{(\Delta\nu)^2}
{2\tau} \right\rangle \right) \right]\right.\nonumber\\
&&\left.+ \nu^2
\left\langle \frac{(\Delta\nu)^2} {2\tau}\right\rangle 
\frac{\partial n}{\partial \nu} \right\},
\label{eq:FP}
\end{eqnarray}
where $n(\nu,t)$ is the photon distribution function, $\langle
\Delta\nu/\tau \rangle$ is the mean change in the photon frequency
$\nu$ per scattering, $\langle (\Delta\nu)^2 / 2\tau \rangle$ is
the mean dispersion per scattering, and $\tau$ is the
scattering optical depth per unit time.

We rewrite this equation in a more familiar notation by defining
\begin{equation}
t_c = (n_e\sigma_T c) t = \frac{t}{\tau},
\end{equation}
\begin{equation}
x = \frac{h\nu}{kT} = \frac{\epsilon}{kT},
\end{equation}
where $\epsilon$ is the photon energy.  We also define
\begin{equation}
\left\langle \frac{\Delta\nu}{\nu} \right\rangle =
\left\langle \frac{\Delta\epsilon}{\epsilon} \right\rangle = \xi,
\label{eq:ksi}
\end{equation}
\begin{equation}
\left\langle \frac{(\Delta\nu)^2}{2\nu^2}\right\rangle = 
\left\langle \frac{(\Delta\epsilon)^2}{2\epsilon^2}\right\rangle = \zeta,
\label{eq:zeta}
\end{equation}
where $\epsilon = h\nu$ is the photon energy.  Then,
equation~(\ref{eq:FP}) becomes
\begin{equation}
\frac{\partial n}{\partial t_c} = \frac{1}{x^2} \frac{\partial}{\partial x}
\left[\zeta x^4 \frac{\partial n}{\partial x} + \left\{ \frac{\partial}{\partial x}
  \left(\zeta x^4\right) - \xi x^3\right\} n(n+1) \right].
\label{eq:FP_xi_zeta}
\end{equation}
If we substitute in this equation the standard low-temperature,
low-frequency expressions, viz., $\xi = (kT/mc^2) (4-x) \equiv
\theta_e (4-x)$, and $\zeta = \theta_e$, we recover the Kompaneets
equation
\begin{equation}
\frac{\partial n}{\partial t_c} = \theta_e\, \frac{1}{x^2} \frac{\partial}{\partial x}
\left[x^4 \left\{ \frac{\partial n}{\partial x} + n(n+1) \right\} \right].
\end{equation}

In many settings in high-energy astrophysics, the electron temperature
is comparable to or even higher than the electron rest mass. In order
to calculate the effects of Compton scattering in such settings, we
would need to include a large number of higher-order terms in the
Fokker-Planck equation~(\ref{eq:FP}) or, even better, employ the full
scattering integral of Comptonization~\citep[see,
  e.g.,][]{suleimanov12}. Such an approach, however, would be very
expensive computationally and would severely impact our ability to run
long simulations of accretion flows. Instead, we use an approximate
method to calculate the effects of Comptonization in high-temperature
flows, which we describe below.

In the limit of high electron temperatures, when a mono-energetic
ensemble of soft photons is scattered by a relativistic distribution
of electrons, the energy distribution of the scattered photons can be
approximated by a log-normal distribution~\citep[see, e.g., Figure~12
  of][]{pozdnyakov83}. This can be understood given the fact that the
energy of each photon exponentially increases after each scattering,
i.e.,
\begin{equation}
  \epsilon^\prime=\epsilon\; e^y\;,
  \label{eq:energy_y}
\end{equation}
where $\epsilon$ and $\epsilon^\prime$ are the photon energies before
and after scattering and $y$ is a variable whose distribution has a
mean of $\xi$ and a dispersion of $2\zeta$. In the limit of very small
energy change per scattering, i.e., when $y\ll 1$,
equation~(\ref{eq:energy_y}) reduces to
$\epsilon^\prime=\epsilon(1+y)$ and the definitions of $\xi$ and
$\zeta$ become identical to those in
equations~(\ref{eq:ksi})--(\ref{eq:zeta}).  If we approximate the
distribution over the values of $y$ by a normal distribution
\begin{equation}
  P(y)dy\propto \exp\left[-\frac{(y-\xi)^2}{2(2\zeta)}\right]\;,
\end{equation}
then we can write for the distribution over the energies of the scattered
photons
\begin{equation}
  I(\epsilon)  = P(y) \left\vert\frac{d\epsilon}{dy}\right\vert^{-1} 
\end{equation}
or, equivalently,
\begin{equation}
  \epsilon I(\epsilon) \propto \exp\left[
    - \frac{\ln(\epsilon/\epsilon_{\rm max})^2}{4\zeta}\right]\;,
\end{equation}
where $\epsilon_{\rm max}=\epsilon\; e^\xi$ is the most likely value for
the energy.

Under this approximation, we can now use the same Fokker-Planck form for
the evolution of the photon distribution function, but with this more
general definition for $\xi$ and $\zeta$. Effectively, we are
integrating out all the higher-order terms in the expansion of the
Fokker-Planck equation, incorporating their effects by modifying the
functional forms of $\xi$ and $\zeta$. In obtaining these
modifications, we are guided by the fact that the photon distribution
function has certain well-defined properties.  In particular, as $t_c
\to \infty$, the photon distribution function has to settle down to a
Bose-Einstein distribution,
\begin{equation}
n(x) = \frac{1}{C e^x - 1},
\end{equation}
which automatically satisfies
\begin{equation}
\frac{\partial n}{\partial x} = -n(n+1).
\end{equation}
Furthermore, in the limit $t_c \to \infty$, the quantity inside the
square brackets $[...]$ in equation (\ref{eq:FP_xi_zeta}), which
represents the flux of photons along the energy coordinate $x$, has to
vanish (thermodynamic equilibrium). Therefore, we have the strong
requirement that
\begin{equation}
  \frac{\partial}{\partial x} (\zeta x^4) - \zeta x^4 - \xi x^3 = 0\;.
  \label{eq:xi_zeta}
\end{equation}
Inserting this into equation~(\ref{eq:FP_xi_zeta}), we obtain
\begin{equation}
\frac{\partial n}{\partial t_c} = \frac{1}{x^2} \frac{\partial}{\partial x}
\left\{\zeta x^4 \left[\frac{\partial n}{\partial x}+ n(n+1) \right]\right\}.
\end{equation}

In the limit of very small photon energy before scattering, $\zeta$
becomes independent of $x$~\citep[see][]{barbosa82} and, therefore, the
condition~(\ref{eq:xi_zeta}) reduces to $\xi=\zeta(4-x)$ and the
Fokker-Planck equation becomes
\begin{equation}
\frac{\partial n}{\partial t_c} = \frac{1}{x^2} \frac{\partial}{\partial x}
\left\{\zeta x^4 \left[\frac{\partial n}{\partial x}+ n(n+1) \right]\right\}.
\end{equation}
In~\citep{sadowski15b} we used the approximate relation 
\begin{equation}
\xi = \frac{(1+3.683\theta_e+4\theta_e^2)}{(1+\theta_e)}\, \theta_e (4-x),
\end{equation}
derived to fit the high temperature behavior of the scattering process
(when $\epsilon \ll kT$), which obeys the requirement that $\xi/(4-x)$
must be independent of $x$. Incorporating this expression into the
Fokker-Planck equation, we obtain
\begin{eqnarray}
  \frac{\partial n}{\partial t_c} &=&
  \frac{(1+3.683\theta_e+4\theta_e^2)}{(1+\theta_e)}\, \theta_e
  \nonumber\\
  &&
  \frac{1}{x^2} \frac{\partial}{\partial x}
\left\{x^4 \left[\frac{\partial n}{\partial x}+ n(n+1) \right]\right\}.
\end{eqnarray}

\bibliography{ms.bib}

\end{document}